\documentclass[prd,twocolumn,superscriptaddress,nofootinbib,amsmath,amssymb]{revtex4-2}
\usepackage{graphicx}
\usepackage{bm}
\usepackage[colorlinks=true,allcolors=blue]{hyperref}
\newcommand{\slashed}[1]{\not\!#1}

\newcommand{\QN}{QMC-N}
\newcommand{\QDa}{QMC-$\Delta$1}
\newcommand{\QDb}{QMC-$\Delta$2}
\newcommand{\DN}{DD-N}
\newcommand{\DDa}{DD-$\Delta$1}
\newcommand{\DDb}{DD-$\Delta$2}

\begin{document}

\title{$\Delta(1232)$ isobars and weak-interaction bulk viscosity in neutron-star merger matter}

\author{Bharat Kumar}
\affiliation{National Institute of Technology Rourkela, Odisha 769008, India}

\date{\today}

\begin{abstract}
Binary neutron star mergers probe nuclear matter at densities of a few times nuclear saturation density and temperatures of tens of MeV, conditions under which heavier baryons such as the $\Delta(1232)$ isobars may appear. Previous calculations of the bulk viscosity of merger matter have considered nucleonic matter, with and without muons, and hyperonic matter, but not $\Delta$ isobars. In this work, we extend these calculations to neutrino-transparent matter in which $\Delta$ isobars are present in strong equilibrium. We derive the relativistic matrix elements for the charged-current $\Delta\to N$ and $\Delta\to\Delta'$ transitions, evaluate all electronic and muonic Urca processes over the full phase space using the nucleon width approximation, and include the in-medium $\Delta\to N\pi$ width with Pauli blocking. The bulk viscosity is obtained from the coupled linear response of the electron and muon fractions about the finite-temperature flavor equilibrium. We find that the $\Delta^-$ takes over the negative charge from the leptons, which raises the proton fraction and opens the nucleonic direct Urca process. Up to about 3~MeV this change in composition, rather than the $\Delta$ reactions themselves, controls flavor equilibration. As a consequence, the resonant maximum of the bulk viscosity moves to lower temperatures for all oscillation frequencies we consider, and kHz density oscillations near resonance are damped several times faster than in nucleonic matter. We also show that, for nucleonic matter, the choice of the equilibrium state about which the rates are linearized shifts the resonance temperature by up to a few MeV at high density.
\end{abstract}

\maketitle

\section{Introduction}
\label{sec:intro}
Neutron stars contain the densest matter in the universe, and binary neutron star mergers in particular probe nuclear matter at densities of a few times nuclear saturation density $n_0$ and temperatures of tens of MeV. In the violent dynamics of a merger we expect high-amplitude density oscillations in the kHz frequency range that drive the matter out of flavor ($\beta$) equilibrium. The weak-interaction (``Urca'') processes that restore equilibrium can then produce strong bulk viscous damping of these oscillations~\cite{Alford2018PRL}. Most \textit{et al.}~\cite{Most2021} post-processed a merger simulation that did not itself include bulk viscosity and found that the bulk viscosity reaches large values in parts of the remnant; their estimates indicated that its effect on the global post-merger dynamics could become significant, whereas during the inspiral it is probably too small to be observable. Simulations that evolve the out-of-equilibrium composition have since shown how a bulk-viscous pressure emerges in the post-merger dynamics~\cite{Most2024}. Whether such effects are observable is still open: Zhu \textit{et al.}~\cite{Zhu2026} estimated that the resulting shift of the post-merger gravitational-wave peak frequency could be detected by third-generation detectors, in optimistic scenarios, only for equations of state with a large slope of the symmetry energy. For a review of bulk viscosity in dense matter, see Ref.~\cite{Harris2024}.

These estimates rely on the microscopic Urca rates, which have been calculated with increasing care for nucleonic matter. At the temperatures above about 1~MeV attained in mergers, the Fermi surface approximation is no longer reliable, and the flavor-equilibrium condition acquires a correction to the cold condition, in which the neutron chemical potential equals the sum of the proton and electron chemical potentials~\cite{AlfordHarris2018,AHHZ2021}. The bulk viscosity must then be calculated by linearizing about the true, finite-temperature equilibrium~\cite{AHZ2024iso}. The nucleon width approximation (NWA)~\cite{AHZ2024nwa} includes the collisional-broadening (``modified Urca'') contribution on the same footing as the direct Urca contribution, and it has been used to study the effect of magnetic fields on the Urca rates~\cite{Tambe2024} and on the bulk viscosity~\cite{Tambe2026}. When muons are present there are separate electronic and muonic Urca processes and hence two relaxation rates~\cite{AHS2023}; Hern\'andez, Manuel and Tolos~\cite{HMT2026} found that this can produce a double-peaked frequency dependence of the bulk viscosity, because the electronic and muonic direct Urca processes open at different densities. Near saturation density, the equation-of-state dependence of the bulk viscosity is determined by the symmetry energy and its slope~\cite{Yang2025}. A common theme of these studies is that the bulk viscosity depends sensitively on the composition of the matter, and in particular on whether the direct Urca process is kinematically allowed.

The composition can change qualitatively if heavier baryons appear. In relativistic mean-field models in which the $\Delta(1232)$ couples somewhat more strongly to the scalar meson than the nucleon does, the $\Delta^-$ appears at two to three times $n_0$, while the maximum mass of neutron stars still exceeds two solar masses~\cite{Drago2014,Cai2015,Kolomeitsev2017,LiSedrakianWeber2018,Raduta2021}. Raduta~\cite{Raduta2021} showed that in cold DDME2 matter the $\Delta^-$ replaces electrons as the negative charge, which raises the proton fraction and opens the nucleonic direct Urca process. The $\Delta$s can also undergo direct Urca processes of their own; their neutrino emissivity has been computed recently~\cite{Yogeesh2026} and included in neutron star cooling simulations~\cite{Yogeesh2026cool}. These studies concern neutrino emission from cold matter, and the response of hot $\Delta$-admixed matter to compression has, to our knowledge, not been studied. For hyperons, the other commonly considered heavy baryons, the strangeness-changing rates and the resulting bulk viscosity have been calculated under merger conditions~\cite{AlfordHaber2021}, and more recent work has addressed their role in the inspiral~\cite{Kantor2026,Storbacka2026}. The $\Delta$s differ from the hyperons in one important respect: strong processes such as $NN\leftrightarrow N\Delta$ keep the $\Delta$ population in equilibrium at all times, so the $\Delta$s do not introduce a new slowly relaxing quantity. Instead, they modify both the thermodynamic response of the matter and the set of weak processes through which the lepton fractions relax. The main thrust of this paper is to explore these two effects.

To do this, we extend the full phase-space NWA calculation of Refs.~\cite{AHZ2024nwa,Tambe2026}, whose nucleonic results we first reproduce, to matter containing the $\Delta$ quartet. We derive the relativistic spin-summed matrix elements for the charged-current $\Delta\to N$ and $\Delta\to\Delta'$ transitions from the Rarita--Schwinger spin sum and include all electronic and muonic channels. We calculate the in-medium $\Delta\to N\pi$ width with Pauli blocking to check whether the populated $\Delta$s are stable against strong decay. We then obtain the bulk viscosity from the coupled linear response of the electron and muon fractions about the finite-temperature flavor equilibrium, and evaluate it as a function of density, temperature and oscillation frequency. In our calculations we assume that the matter is transparent to neutrinos, which is a reasonable approximation below about 10~MeV~\cite{AHZ2024iso,Tambe2026}.

This paper is structured as follows. In Sec.~\ref{sec:eos} we describe the models of $\Delta$-admixed nuclear matter. In Sec.~\ref{sec:rates} we derive the matrix elements of the $\Delta$ transitions and the in-medium $\Delta$ width, and in Sec.~\ref{sec:bulk} we present the two-variable linear response. Section~\ref{sec:valid} summarizes our numerical checks, Sec.~\ref{sec:results} presents our results, and in Sec.~\ref{sec:concl} we discuss their implications. Standard ingredients that we use without modification, namely the mean-field equations, the phase-space reduction, the NWA, the nucleonic matrix element, and the one-variable and damping-time relations, are collected in Appendices~\ref{app:rmf}--\ref{app:bv}, and all parameter values are listed in Tables~\ref{tab:param}--\ref{tab:inputs}.

\section{Matter with $\Delta$ isobars}
\label{sec:eos}
We use natural units, in which the reduced Planck constant, the speed of light and the Boltzmann constant are set to one, and the metric signature $(+,-,-,-)$ throughout this paper, and we convert to CGS units only for the bulk viscosity and the damping time. Baryon densities are given in units of the reference density $n_0$ of Table~\ref{tab:inputs}.

\subsection{Models}
We describe nuclear matter by relativistic mean-field theories containing the nucleon doublet and the $\Delta(1232)$ quartet, with the $\Delta$s described by Rarita--Schwinger vector-spinors~\cite{RS1941}. The baryons interact through $\sigma$, $\omega$ and $\rho$ exchange, which we treat in the mean-field approximation. Each $\Delta$ is then a spin-$\tfrac32$ quasiparticle with four spin states, an effective mass and a vector self-energy, and its meson couplings are specified as fractions of the nucleon couplings. The Lagrangian, the field equations and the finite-temperature thermodynamics are given in Appendix~\ref{app:rmf}.

One of the most important features influencing flavor equilibration is the direct Urca threshold. To isolate the effect of the $\Delta$s, we use two nucleonic models in which the nucleonic direct Urca process is kinematically forbidden in cold $\beta$-equilibrated matter up to eight times $n_0$. QMC-RMF3~\cite{ABHT2022,ABHT2023} has constant couplings and an $\omega\rho$ mixing term; it is the model without a direct Urca threshold used in Refs.~\cite{AHZ2024iso,Tambe2026}. DDME2~\cite{DDME2} has density-dependent couplings, and the absence of a direct Urca threshold agrees with Refs.~\cite{Raduta2021,AHS2023}. Any opening of the direct Urca process in our calculations can therefore be attributed to the $\Delta$s. The parameters and saturation properties of both models are listed in Table~\ref{tab:param}. The saturation properties obtained with our solver agree with Ref.~\cite{ABHT2022} for QMC-RMF3 and, for DDME2, with the symmetry energy and slope quoted in Ref.~\cite{Raduta2021}.

\begin{table*}[t]
\caption{Nucleonic parameters and saturation properties. QMC-RMF3: constant couplings of Ref.~\cite{ABHT2022}, Table~I; the parameters $b$, $c$, $b_1$ of that reference translate into $\kappa=2bM_N$, $\lambda=6c$ and $\Lambda_v=b_1/g_\omega^2$ of Eq.~\eqref{eq:Uqmc}. DDME2: couplings at saturation and the density-dependence parameters of Eq.~\eqref{eq:dd}~\cite{DDME2}. The isospin factor $\theta_b$ of the $\rho$ coupling in Eq.~\eqref{eq:Ub} differs between the two models. Masses in MeV. Lower block: saturation density (fm$^{-3}$), binding energy per baryon, incompressibility, symmetry energy and its slope (MeV), and Dirac effective mass, computed with our solver; for QMC-RMF3 the symmetry energy and slope are from the difference between pure neutron and symmetric matter, for DDME2 from the quadratic term.}
\label{tab:param}
\begin{ruledtabular}
\begin{tabular}{lll}
 & QMC-RMF3 & DDME2 \\ \hline
$M_N$ & 939.0 & 939.0 \\
$m_\sigma,\ m_\omega,\ m_\rho$ & 491.5, 782.5, 763.0 & 550.1238, 783.0, 763.0 \\
$g_\sigma,\ g_\omega,\ g_\rho$ & 8.32, 9.76, 11.02 & 10.5396, 13.0189, 3.6836 (at $n_{\rm sat}$) \\
$\theta_b$ in Eq.~\eqref{eq:Ub} & $t_{3b}$ & $2t_{3b}$ \\
$b$, $c$, $b_1$ (Ref.~\cite{ABHT2022}) & 0.0063, $-0.006$, 5.87 & -- \\
$\kappa$ (MeV), $\lambda$, $\Lambda_v$ & 11.83, $-0.036$, 0.0616 & -- \\
$a_\sigma,b_\sigma,c_\sigma,d_\sigma$ & -- & 1.3881, 1.0943, 1.7057, 0.4421 \\
$a_\omega,b_\omega,c_\omega,d_\omega$ & -- & 1.3892, 0.9240, 1.4620, 0.4775 \\
$a_\rho$ & -- & 0.5647 \\
$n_{\rm sat}$ in Eq.~\eqref{eq:dd} & -- & 0.152 \\ \hline
$n_{\rm sat}$ & 0.1569 & 0.1519 \\
$E/A$ & $-16.06$ & $-16.14$ \\
$K$ & 229.8 & 250.9 \\
$J,\ L$ & 33.55, 49.0 & 32.30, 51.3 \\
$M^*/M_N$ & 0.732 & 0.572 \\
\end{tabular}
\end{ruledtabular}
\end{table*}

\subsection{Strong and weak equilibrium}
On strong-interaction timescales the baryon composition equilibrates through processes such as $NN\leftrightarrow N\Delta$, which are many orders of magnitude faster than the weak processes we consider. At fixed baryon density and charge fraction the baryon chemical potentials therefore obey
\begin{equation}
\mu_b = \mu_B + q_b\mu_Q ,\qquad \mu_B\equiv\mu_n,\quad \mu_Q\equiv\mu_p-\mu_n ,
\label{eq:strong}
\end{equation}
that is,
\begin{align}
\mu_{\Delta^-}&=2\mu_n-\mu_p,\qquad \mu_{\Delta^0}=\mu_n, \nonumber\\
\mu_{\Delta^+}&=\mu_p,\qquad\qquad\ \ \mu_{\Delta^{++}}=2\mu_p-\mu_n .
\label{eq:strong2}
\end{align}
For given baryon density, charge fraction and temperature we solved the conditions of baryon number and charge,
\begin{equation}
\sum_b n_b=n_B,\qquad \sum_b q_bn_b=Y_Qn_B=n_e+n_\mu ,
\label{eq:cons}
\end{equation}
together with the field equations, for the unknowns $\mu_B$, $\mu_Q$, $\Phi$, $W$, $B$ (and $\Sigma_R$ for DDME2) by a multidimensional Newton method, to a relative residual below $10^{-8}$.

On longer timescales the weak interactions change the lepton numbers. In neutrino-transparent matter the electron and muon fractions $Y_e$ and $Y_\mu$ (lepton densities per baryon) change only through Urca processes, and, as in Ref.~\cite{AHS2023}, we assume that purely leptonic conversion between muons and electrons is frozen. A state out of weak equilibrium is then specified by the baryon density, the temperature and the two lepton fractions, and the baryon sector depends only on their sum $Y_Q$. The departures from cold $\beta$ equilibrium are
\begin{equation}
\delta\mu_e = \mu_n-\mu_p-\mu_e ,\qquad \delta\mu_\mu=\mu_n-\mu_p-\mu_\mu .
\label{eq:dmu}
\end{equation}
Because of Eq.~\eqref{eq:strong}, every charged-current process that converts a baryon into one with one unit more charge involves the same chemical-potential difference between initial and final baryon, $\mu_n-\mu_p$. This means that a single departure $\delta\mu_l$ drives all the processes involving a given lepton.

\subsection{$\Delta$ coupling sets}
For each model we use two $\Delta$ coupling sets within the ranges discussed in Refs.~\cite{Drago2014,Kolomeitsev2017,LiSedrakianWeber2018,Raduta2021}, which differ only in the scalar coupling ratio. The six resulting equations of state are listed in Table~\ref{tab:sets}, together with the $\Delta$ onset densities and stellar properties, and we refer to them by the names given there. All four $\Delta$ sets support maximum masses of at least two solar masses, and our nucleonic maximum masses agree with the published values~\cite{ABHT2022,Raduta2021}; the radii are less reliable because we use a polytropic crust. A scalar ratio of 1.3 produces a pressure that decreases with density in QMC-RMF3, so we exclude it. All four remaining sets are stable against spinodal decomposition between one and six times $n_0$ (Appendix~\ref{app:rmf}). This is consistent with Ref.~\cite{Raduta2021}, which found $\Delta$-driven instabilities in DDME2 only for a larger excess of the scalar over the vector coupling ratio than in our sets.

\begin{table*}[t]
\caption{The six equations of state. Left block: $\Delta$ coupling ratios of Eq.~\eqref{eq:xcoup}. Middle block: onset densities of the $\Delta$ species and the cold nucleonic direct Urca threshold in $\beta$-equilibrated matter with muons at 1~MeV. Right block: maximum mass, its radius and central density, and the radius and central density of a $1.4\,M_\odot$ star (cold, polytropic crust). Densities in units of $n_0$; a dash means not reached below six times $n_0$ (onsets) or eight times $n_0$ (direct Urca). For DD-$\Delta$1 the direct Urca triangle is open only between the two densities given.}
\label{tab:sets}
\begin{ruledtabular}
\begin{tabular}{lcccccccccc}
Name & $x_\sigma$ & $x_\omega$ & $x_\rho$ & $n_{\Delta^-}$ & $n_{\Delta^0}$ & $n_{\Delta^+}$ & $n_{\rm dU}$ & $M_{\max}$ [$M_\odot$] & $R$, $n_c$ at $M_{\max}$ [km, $n_0$] & $R$, $n_c$ at $1.4\,M_\odot$ \\ \hline
\QN  & -- & -- & -- & -- & -- & -- & -- & 2.147 & 10.64, 6.67 & 12.00, 2.92 \\
\QDa & 1.1 & 1.1 & 1.0 & 2.30 & 4.70 & -- & 3.25 & 2.028 & 9.99, 7.56 & 11.67, 3.53 \\
\QDb & 1.2 & 1.1 & 1.0 & 1.95 & 3.45 & 4.30 & 2.70 & 2.012 & 9.66, 7.92 & 11.04, 4.00 \\
\DN  & -- & -- & -- & -- & -- & -- & -- & 2.489 & 12.05, 5.05 & 13.01, 2.11 \\
\DDa & 1.1 & 1.1 & 1.0 & 1.80 & 4.40 & -- & 3.15--4.75 & 2.392 & 11.47, 5.64 & 12.86, 2.37 \\
\DDb & 1.2 & 1.1 & 1.0 & 1.45 & 2.75 & 3.95 & 2.10 & 2.379 & 11.22, 5.77 & 12.25, 2.63 \\
\end{tabular}
\end{ruledtabular}
\end{table*}

\section{Weak rates with $\Delta$ isobars}
\label{sec:rates}
\subsection{Channels and rates}
For each charged lepton, electron or muon, we included the decay and capture branches of
\begin{align}
&n\to p\,l\,\bar\nu_l , && \Delta^-\to n\,l\,\bar\nu_l , && \Delta^0\to p\,l\,\bar\nu_l , \nonumber\\
&\Delta^-\to\Delta^0 l\,\bar\nu_l , && \Delta^0\to\Delta^+ l\,\bar\nu_l , && \Delta^+\to\Delta^{++} l\,\bar\nu_l ,
\label{eq:channels}
\end{align}
together with the inverse lepton-capture reactions. The net lepton production rate per unit volume is
\begin{equation}
\Gamma_l(n_B,T,Y_e,Y_\mu) = \sum_{\rm ch}\big[\Gamma^{\rm dec}_{{\rm ch},l} - \Gamma^{\rm cap}_{{\rm ch},l}\big] ,
\label{eq:Gnet}
\end{equation}
and we defined weak equilibrium at temperature $T$ by the vanishing of both net rates,
\begin{equation}
\Gamma_e=\Gamma_\mu=0 ,
\label{eq:thermeq}
\end{equation}
which differs from the vanishing of the chemical imbalances by the thermal shift discussed in Ref.~\cite{AHHZ2021}.

We calculate each rate as the full phase-space integral of Ref.~\cite{AHHZ2021}, reduced exactly to five dimensions, and average it over Breit--Wigner spectral functions of both baryons following the NWA of Ref.~\cite{AHZ2024nwa}. We have generalized both steps to unequal baryon masses, arbitrary vector potentials and spin-$\tfrac32$ baryons, so that the $\Delta$s enter only through their dispersion relations, their spectral widths and the matrix elements. The reduction, the NWA and the nucleonic matrix element are given in Appendix~\ref{app:rates}. For the collisional width of the $\Delta$ we take the nucleon value and add the in-medium $\Delta\to N\pi$ width of Sec.~\ref{sec:width}; Appendix~\ref{app:rates} gives the reasons for this choice, and Fig.~\ref{fig:sens} its effect. What remains is to specify the matrix elements of the $\Delta$ transitions, which we derive below.

\subsection{$\Delta$ transition matrix elements}
\label{sec:mel}
The effective charged-current interaction, the lepton current and the spin-summed lepton tensor are given in Appendix~\ref{app:rates}; the hadronic tensors derived below are contracted with this lepton tensor.

\textit{$\Delta\to N$.} The leading term of the $N\Delta$ axial transition current at low momentum transfer and the resulting amplitude are~\cite{HNV2007}
\begin{equation}
\langle N|A^\mu|\Delta\rangle = C_5^A\,\bar u_N\,u^\mu_\Delta,\qquad
\mathcal M=\frac{G}{\sqrt2}\,\mathcal I\, C_5^A\,\bar u_N u^\mu_\Delta\, l_\mu ;
\label{eq:JDN}
\end{equation}
the vector form factors and the other axial form factors are suppressed by powers of the momentum transfer over the $\Delta$ mass, and we neglected them. In our conditions the momentum transfer is of the order of the electron Fermi momentum, about 100~MeV or less, so this ratio is below about 0.1. The isospin factors $|\mathcal I|^2$ and the value of $C_5^A$ and its variation are given in Table~\ref{tab:inputs}; the isospin factors are normalized to one for the nucleon. The RS spin sum is
\begin{align}
\sum_s u^\mu_\Delta\bar u^\nu_\Delta = -(\slashed p_\Delta+M)\Big[&g^{\mu\nu}-\tfrac13\gamma^\mu\gamma^\nu-\tfrac{2p_\Delta^\mu p_\Delta^\nu}{3M^2} \nonumber\\
&+\tfrac{p_\Delta^\mu\gamma^\nu-p_\Delta^\nu\gamma^\mu}{3M}\Big] ,
\label{eq:RSsum}
\end{align}
with $M$ the $\Delta$ effective mass, and the hadronic tensor, with $m$ the nucleon effective mass, is
\begin{equation}
H^{\mu\nu}=\mathrm{Tr}\Big[(\slashed p_N+m)\sum_s u^\mu_\Delta\bar u^\nu_\Delta\Big] .
\label{eq:Hdef}
\end{equation}
It follows term by term from the trace identities
\begin{align}
&\mathrm{Tr}[\slashed a\slashed b]=4\,a\!\cdot\!b,\qquad \mathrm{Tr}[\gamma^\mu\gamma^\nu]=4g^{\mu\nu}, \nonumber\\
&\mathrm{Tr}[\slashed a\slashed b\gamma^\mu\gamma^\nu]=4[(a\!\cdot\!b)g^{\mu\nu}-a^\mu b^\nu+a^\nu b^\mu] ,
\label{eq:trid}
\end{align}
the on-shell condition for the $\Delta$ and the vanishing of traces of odd numbers of $\gamma$ matrices. The four terms of Eq.~\eqref{eq:RSsum} give
\begin{align}
H^{\mu\nu}_{(1)} &\equiv -\mathrm{Tr}[(\slashed p_N+m)(\slashed p_\Delta+M)]\,g^{\mu\nu} \nonumber\\
&= -4(p_N\!\cdot\!p_\Delta+mM)\,g^{\mu\nu}, \nonumber\\
H^{\mu\nu}_{(2)} &\equiv \tfrac13\mathrm{Tr}[(\slashed p_N+m)(\slashed p_\Delta+M)\gamma^\mu\gamma^\nu] \nonumber\\
&= \tfrac43(p_N\!\cdot\!p_\Delta+mM)\,g^{\mu\nu}+\tfrac43\big(p_\Delta^\mu p_N^\nu-p_\Delta^\nu p_N^\mu\big), \nonumber\\
H^{\mu\nu}_{(3)} &\equiv \tfrac{2p^\mu_\Delta p^\nu_\Delta}{3M^2}\mathrm{Tr}[(\slashed p_N+m)(\slashed p_\Delta+M)] \nonumber\\
&= \tfrac{8}{3M^2}(p_N\!\cdot\!p_\Delta+mM)\,p^\mu_\Delta p^\nu_\Delta, \nonumber\\
H^{\mu\nu}_{(4)} &\equiv -\tfrac{1}{3M}\mathrm{Tr}[(\slashed p_N+m)(\slashed p_\Delta+M)(p_\Delta^\mu\gamma^\nu-p_\Delta^\nu\gamma^\mu)] \nonumber\\
&= -\tfrac{4}{3}\big(p_\Delta^\mu p_N^\nu - p_\Delta^\nu p_N^\mu\big) .
\label{eq:traces}
\end{align}
In the fourth term the contributions proportional to the nucleon mass and to the product of two $\Delta$ momenta cancel between the two orderings. The antisymmetric pieces of the second and fourth terms cancel, and the sum is
\begin{equation}
H^{\mu\nu} = -\frac83\,(p_N\!\cdot\!p_\Delta + mM)\Big[g^{\mu\nu}-\frac{p_\Delta^\mu p_\Delta^\nu}{M^2}\Big] .
\label{eq:Hdn}
\end{equation}
The hadronic tensor is thus symmetric and transverse to the $\Delta$ momentum, as required by the RS constraints, and the antisymmetric part of the lepton tensor does not contribute. The two contractions needed are
\begin{align}
g^{\mu\nu}L_{\mu\nu}&=-16\,k_l\!\cdot\!k_\nu, \nonumber\\
p^\mu p^\nu L_{\mu\nu}&=8\big[2(p\!\cdot\!k_l)(p\!\cdot\!k_\nu)-M^2\,k_l\!\cdot\!k_\nu\big] ,
\label{eq:contr}
\end{align}
and they give
\begin{align}
\overline\sum|\mathcal M|^2 ={}& \frac{32}{3}G^2|\mathcal I|^2(C_5^A)^2\,(p_N\!\cdot\!p_\Delta+m^*_Nm^*_\Delta) \nonumber\\
&\times\Big[k_l\!\cdot\!k_\nu+\frac{2(p_\Delta\!\cdot\!k_l)(p_\Delta\!\cdot\!k_\nu)}{m^{*2}_\Delta}\Big] .
\label{eq:MDN}
\end{align}
For baryons at rest the angular average of the bracket is three times the product of the lepton energies, so
\begin{equation}
\overline\sum|\mathcal M|^2\to64G^2|\mathcal I|^2(C_5^A)^2\,m^*_Nm^*_\Delta\,E_lE_\nu .
\label{eq:MDNnr}
\end{equation}
Per initial spin state this is the nucleon value of Eq.~\eqref{eq:MNnr} multiplied by the isospin factor and the squared transition coupling, divided by the nucleon Gamow--Teller factor $1+3g_A^2$. We verified Eq.~\eqref{eq:Hdn} with the symbolic program FORM, and Eq.~\eqref{eq:MDN} against an explicit sum over RS spinors built from Clebsch--Gordan products of Dirac spinors and polarization vectors, at random off-axis relativistic kinematics; the ratio was one to ten digits.

\textit{$\Delta\to\Delta'$.} For transitions inside the quartet we used the RS current and its isospin factor
\begin{align}
J^\mu&=-\bar u_{\Delta'\alpha}\gamma^\mu(1-g_{A\Delta}\gamma_5)u^\alpha_\Delta, \nonumber\\
|\langle\Delta'|T_+|\Delta\rangle|^2&=I(I+1)-t_3(t_3+1) ,
\label{eq:JDD}
\end{align}
which gives the factors 3, 4 and 3 of Table~\ref{tab:inputs} for the three transitions. The axial coupling was fixed by the quark model. In the state of maximal spin projection all quark spins are aligned, so the quark-model Gamow--Teller operator reduces to the isospin raising operator, while the RS axial current measures only the Dirac-spin part of the spin-$\tfrac32$ state, which by the projection theorem is two thirds of the total spin projection:
\begin{equation}
\Big\langle\Delta'\Big|\sum_q\sigma_z^q t_+^q\Big|\Delta\Big\rangle=\langle T_+\rangle,\qquad \langle 2s_z\rangle=\tfrac23S_z .
\label{eq:QM}
\end{equation}
Normalizing to the quark-model nucleon value of $5/3$ then gives
\begin{equation}
g_{A\Delta}=\tfrac35\,g_A ,
\label{eq:gAD}
\end{equation}
with the numerical value of Table~\ref{tab:inputs}. Since $\langle\Delta|T_3|\Delta\rangle=\tfrac32$ for the $\Delta^{++}$, this coupling corresponds to the $\Delta^{++}$ axial charge $g_1^{\Delta^{++}}(0)=3g_{A\Delta}=\tfrac95g_A\approx2.27$ in the normalization of Refs.~\cite{JiangTiburzi2008,Jun2020}. Chiral perturbation theory and a relativistic constituent-quark model give about 2.25, the chiral quark-soliton model 2.13, and lattice QCD at pion masses of 213--432~MeV gives 1.7--2.0, as compiled in Ref.~\cite{Jun2020}. The lattice range corresponds to $g_{A\Delta}\approx0.57$--0.66, which lies inside the range $0\le g_{A\Delta}\le g_A$ of our sensitivity test (Sec.~\ref{sec:results}). The spin-summed hadron--lepton contraction, computed with FORM, is given in Appendix~\ref{app:DD}; it reproduced explicit RS spinor sums to $10^{-10}$. For both $\Delta$s at rest it reduces to
\begin{equation}
\overline\sum|\mathcal M|^2 \to \frac{64}{3}G^2|\langle T_+\rangle|^2(3+5g_{A\Delta}^2)\,m^*_\Delta m^*_{\Delta'}E_lE_\nu .
\label{eq:MDDnr}
\end{equation}
Per initial state the Fermi strength is therefore the same as for the nucleon, sixteen times the isospin factor, and the Gamow--Teller strength is the nucleon one multiplied by the mean squared Dirac spin of the spin-$\tfrac32$ state relative to the nucleon, which is five ninths, and by the squared ratio of the axial couplings.

\subsection{In-medium $\Delta$ width}
\label{sec:width}
In vacuum the $\Delta$ decays strongly to $N\pi$. In the medium, energy conservation for the decay of a $\Delta$ of momentum $\bm p$ into a mean-field nucleon and a vacuum pion, the available four-momentum and the decay momentum $q$ in its rest frame are
\begin{align}
&E^*_\Delta(\bm p)+U_\Delta-U_N=E^*_N+\omega_\pi , \nonumber\\
&\mathcal P=(E^*_\Delta+U_\Delta-U_N,\bm p), \nonumber\\
&\sqrt{q^2+m_N^{*2}}+\sqrt{q^2+m_\pi^2}=\sqrt s ,\qquad s=\mathcal P^2 .
\label{eq:wkin}
\end{align}
We used the $P$-wave form
\begin{align}
\Gamma^{\rm med}_\Delta(\bm p) ={}& \sum_{N}\mathrm{BR}_N\,\Gamma_0\Big(\frac{q}{q_0}\Big)^3\frac{m_N^*/(E^*_N+\omega_\pi)}{M_N/M_\Delta} \nonumber\\
&\times\frac{\sqrt s}{\mathcal P^0}\,\big\langle1-f_N(E_N^{\rm lab})\big\rangle ,
\label{eq:Gmed}
\end{align}
normalized to the vacuum Breit--Wigner width at the Breit--Wigner mass~\cite{PDG2024}; these values, the corresponding vacuum decay momentum and the isospin branching ratios are listed in Table~\ref{tab:inputs}. The Pauli factor is averaged over the decay angle after boosting the nucleon to the matter frame, and the ratio of the invariant mass to the energy is the time dilation. The width entering the NWA is the Fermi-surface average with weight proportional to the squared momentum times the product of the occupation and its complement. For a $\Delta$ at rest the formula reproduces the rest-frame value, and in vacuum it returns a decay momentum of 226.5~MeV and a width of 116.3~MeV. Pion self-energies and Bose enhancement were neglected.

\begin{table}[t]
\caption{Weak-interaction, $\Delta$ and numerical inputs. Masses and energies in MeV.}
\label{tab:inputs}
\begin{ruledtabular}
\begin{tabular}{p{4.6cm}p{3.3cm}}
Quantity & Value \\ \hline
Reference density $n_0$ & 0.16~fm$^{-3}$ \\
Fermi constant $G_F$ & $1.166\times10^{-11}$~MeV$^{-2}$ \\
Cabibbo angle $\theta_C$ & $13.04^\circ$ \\
Nucleon axial coupling $g_A$ & 1.26 \\
$N\Delta$ axial coupling $C_5^A$ & 1.2 (varied to 0.87) \\
$\Delta\Delta'$ axial coupling $g_{A\Delta}$, Eq.~\eqref{eq:gAD} & 0.756 (varied to 0) \\
Isospin factor $|\mathcal I|^2$: $\Delta^-\to n$, $\Delta^0\to p$ & 3, 1 \\
Isospin factor $|\langle T_+\rangle|^2$: $\Delta^-\to\Delta^0$, $\Delta^0\to\Delta^+$, $\Delta^+\to\Delta^{++}$ & 3, 4, 3 \\
Lepton masses $m_e$, $m_\mu$ & 0.511, 105.658 \\
$\Delta$ mass $M_\Delta$ (PDG estimate~\cite{PDG2024}) & 1232 (1230--1234) \\
Vacuum width $\Gamma_0$ (PDG estimate~\cite{PDG2024}) & 117 (114--120) \\
Vacuum decay momentum $q_0$; pion mass $m_\pi$ & 227; 139.57 \\
$N\pi$ branching: $\Delta^-$; $\Delta^0$ ($p$, $n$); $\Delta^+$ ($p$, $n$); $\Delta^{++}$ & 1; $\tfrac13$, $\tfrac23$; $\tfrac23$, $\tfrac13$; 1 \\
NWA width scale $T_W$, Eq.~\eqref{eq:WN} & 5~MeV \\
Extra $\Delta$ width in sensitivity test & 10 \\
Oscillation frequency $f$ & 1~kHz; 0.5--5~kHz in Fig.~\ref{fig:fmaps} \\
Damping-time contour in Fig.~\ref{fig:fmaps} & 20~ms \\ \hline
Quasi-Monte Carlo points per rate & $2^{17}$ \\
Energy window around Fermi energies & $\pm16T$ (plus $|\delta\mu_l|$) \\
Gauss--Chebyshev nodes in $z_l$ & 6 \\
Finite-difference steps in $Y_l$ and $n_B$ & $2\times10^{-3}\,Y_l$, $10^{-3}\,n_B$ \\
Equilibrium tolerance (net/gross rate) & $10^{-5}$ \\
Density grid & 1--6 $n_0$, step $0.5\,n_0$; $0.25\,n_0$ for nucleonic matter at 1, 2, 3~MeV \\
Grid of Fig.~\ref{fig:fmaps} (isothermal) & 1--6 $n_0$, step $0.25\,n_0$; $T=$ 1, 1.5, 2--4.5 (step 0.25), 5, 5.5, 6, 7, 8, 10~MeV \\
Temperature grid & 1, 2, 2.5, 3, 3.5, 4, 4.5, 5, 7, 10; also 1.5, 2.25 for the $\Delta2$ sets \\
\end{tabular}
\end{ruledtabular}
\end{table}

\section{Bulk viscosity with two slow variables}
\label{sec:bulk}
Because strong processes keep the $\Delta$s in equilibrium, they do not introduce a new slow variable, but they enter every ingredient of the response. The pressure and its derivatives are those of matter in strong equilibrium at each perturbed state, and the relaxation matrix contains all the processes of Eq.~\eqref{eq:channels}. We therefore formulate the response directly for the two lepton fractions, generalizing the one-variable treatment of Ref.~\cite{AHZ2024iso}.

\subsection{Coupled response of electrons and muons}
Consider a fluid element undergoing a small periodic compression with angular frequency $\omega$ and fluid velocity $\bm v$,
\begin{equation}
n_B(t)=\bar n_B+\delta n\,e^{i\omega t} .
\label{eq:comp}
\end{equation}
Baryon and lepton number conservation give
\begin{equation}
\dot n_B+n_B\nabla\!\cdot\!\bm v=0,\qquad \dot n_l+n_l\nabla\!\cdot\!\bm v=\Gamma_l ,
\label{eq:conserv}
\end{equation}
hence
\begin{equation}
n_B\frac{dY_l}{dt} = \Gamma_l(n_B,T,Y_e,Y_\mu) ,\qquad l=e,\mu .
\label{eq:cont}
\end{equation}
We linearized about the weak equilibrium of Eq.~\eqref{eq:thermeq}, writing
\begin{equation}
Y_l=\bar Y_l+\delta Y_l\,e^{i\omega t} ,
\label{eq:Ypert}
\end{equation}
and obtained
\begin{equation}
i\omega\,\delta Y_a = -\sum_b\mathcal G_{ab}\,\delta Y_b + \frac{\gamma_{B,a}}{\bar n_B}\,\delta n ,
\label{eq:lin}
\end{equation}
with the relaxation matrix and the density response
\begin{equation}
\mathcal G_{ab}=-\frac{1}{n_B}\frac{\partial\Gamma_a}{\partial Y_b}\Big|_{n_B,Y_{c\ne b}} ,\qquad
\gamma_{B,a}=\frac{\partial\Gamma_a}{\partial n_B}\Big|_{Y_e,Y_\mu} .
\label{eq:Gmat}
\end{equation}
The solution is
\begin{equation}
\delta\bm Y=(i\omega\mathbb 1+\mathcal G)^{-1}\bm\gamma_B\,\frac{\delta n}{\bar n_B} .
\label{eq:dYsol}
\end{equation}
The departure of the pressure from the value it would have at frozen composition, and the bulk viscous pressure that defines the bulk viscosity, are
\begin{align}
\delta P_{\rm neq}&=\sum_a\frac{\partial P}{\partial Y_a}\,\delta Y_a\equiv X(\omega)\,\frac{\delta n}{\bar n_B}, \nonumber\\
\Pi&=-\zeta\,\nabla\!\cdot\!\bm v=\zeta\,i\omega\,\frac{\delta n}{\bar n_B} ,
\label{eq:Pi}
\end{align}
the latter being the component of the former in phase with the compression rate. With
\begin{equation}
X(\omega)=\sum_{a,b}\frac{\partial P}{\partial Y_a}\big[(i\omega\mathbb1+\mathcal G)^{-1}\big]_{ab}\gamma_{B,b} ,
\label{eq:X}
\end{equation}
the dissipative part is the imaginary part of $X$, and
\begin{equation}
\zeta(\omega) = \frac{\mathrm{Im}\,X(\omega)}{\omega} .
\label{eq:zeta2}
\end{equation}
For a single slow variable, without muons, Eq.~\eqref{eq:zeta2} reduces to
\begin{equation}
\zeta = -\frac{\partial P}{\partial Y_e}\,\frac{\gamma_B}{\gamma^2+\omega^2} ,\qquad \gamma=\mathcal G_{ee} ,
\label{eq:zeta1}
\end{equation}
the form of Ref.~\cite{AHZ2024iso} with the isospin fraction replaced by the electron fraction. The eigenvalues of $\mathcal G$ are the two relaxation rates of the coupled electron--muon system, and we call the inverse of the diagonal elements the electron and muon relaxation times,
\begin{equation}
\tau_e=1/\mathcal G_{ee},\qquad \tau_\mu=1/\mathcal G_{\mu\mu} .
\label{eq:tau}
\end{equation}
In the presence of $\Delta$s the pressure derivatives and the density response are evaluated with the baryon composition in strong equilibrium at every state, so the response of the $\Delta$ populations to compression is included at the level of the susceptibilities.

The derivatives were taken by central differences, with the steps of Table~\ref{tab:inputs}, using the same quasi-random point set for every rate evaluation (common random numbers), so that the finite differences of the rates carry no noise from independent samples. As a check of the linear-response machinery we verified, in matter with $\Delta$s, the Maxwell-type identity
\begin{equation}
\frac{1}{n_B}\frac{\partial P}{\partial Y_e}\Big|_{n_B,T} = n_B\frac{\partial(\mu_Q+\mu_e)}{\partial n_B}\Big|_{Y_e,T} ,
\label{eq:identity}
\end{equation}
which follows from the Gibbs--Duhem relation
\begin{equation}
dP=s\,dT+\sum_in_i\,d\mu_i
\label{eq:GD}
\end{equation}
and the strong-equilibrium relations, to $3\times10^{-8}$.

\subsection{Oscillation type, damping time and resonance}
We evaluate Eq.~\eqref{eq:Gmat} both at fixed temperature (isothermal oscillations) and at fixed entropy per baryon (adiabatic oscillations), and convert the bulk viscosity into the damping time of a density oscillation, Eq.~\eqref{eq:tdamp}. These standard relations, together with the resonance form that we use to interpret the frequency dependence, are given in Appendix~\ref{app:bv}.

\section{Validation}
\label{sec:valid}
We first validated our rate code by reproducing the nucleonic results of Ref.~\cite{Tambe2026} (and, through them, Refs.~\cite{AHZ2024iso,AHZ2024nwa}) for IUF and QMC-RMF3, with and without magnetic fields. Over the 88 published curves we compared with, the median deviations in the rates, the relaxation time and the bulk viscosity are a few hundredths of a dex. This comparison used the constant matrix element of Refs.~\cite{Tambe2024,Tambe2026}. With the full matrix element of Eq.~\eqref{eq:MN} the rates are lower by a factor of two to three, and all the results below use the full matrix elements.

We then tested the extension to $\Delta$s in several ways. With the $\Delta$s removed, both the equation of state and the full bulk-viscosity calculation reproduce the nucleonic code to numerical precision. The generalized rate kernel reproduces the nucleon NWA kernel, and for the $\Delta\to N$ processes the full phase-space rate approaches the Fermi surface limit of the same matrix element in degenerate kinematics. The matrix elements of Eqs.~\eqref{eq:MDN} and \eqref{eq:HDD} agree with explicit Rarita--Schwinger spinor sums to ten digits, and the two-variable response reduces to the one-variable result when the muon variable is removed. In matter with $\Delta$s the Gibbs--Duhem relation and the thermodynamic identity of Eq.~\eqref{eq:identity} are satisfied to better than $10^{-7}$. We accept a flavor-equilibrium state only when the net rates are below the tolerance of Table~\ref{tab:inputs} relative to the gross rates. This criterion is not met at one grid point (\QDa\ at $n_0$ and 10~MeV), where the net rate is a small difference between large decay and capture rates, and we exclude that point.

\section{Results}
\label{sec:results}
\subsection{Composition and the opening of direct Urca}
\begin{figure*}[t]
\includegraphics[width=\textwidth]{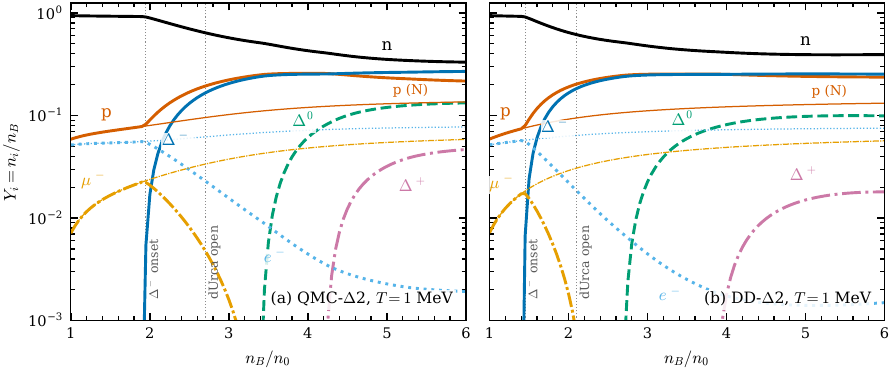}
\caption{Particle fractions of $\beta$-equilibrated matter (equal electron and muon chemical potentials, both equal to the neutron--proton difference) at 1~MeV for (a) \QDb\ and (b) \DDb. Thick lines, labeled on the curves: matter with $\Delta$s. Thin lines of the same color and style: proton, electron and muon fractions of the nucleonic sets \QN\ and \DN, with the proton line marked p (N). Vertical dotted lines mark the $\Delta^-$ onset and the density above which the cold nucleonic direct Urca triangle is open. The $\Delta^-$ replaces electrons and muons as the negative charge and doubles the proton fraction, which opens the direct Urca triangle.}
\label{fig:comp}
\end{figure*}

In Fig.~\ref{fig:comp} we show the composition of $\beta$-equilibrated matter for \QDb\ and \DDb. Once the $\Delta^-$ threshold is crossed, it is energetically cheaper to carry negative charge in a heavy baryon near the bottom of a shallow Fermi sea than in a relativistic electron at the top of a deep one. The $\Delta^-$ therefore takes over the role of the leptons in maintaining charge neutrality: the electron fraction drops by more than an order of magnitude, the muons disappear, and the proton fraction roughly doubles. With the larger proton Fermi momentum, the neutron Fermi momentum no longer exceeds the sum of the proton and electron Fermi momenta, so momentum conservation allows the nucleonic direct Urca process in all four $\Delta$ sets (Table~\ref{tab:sets}), as Raduta found for cold DDME2~\cite{Raduta2021}. For our couplings the corresponding conditions for the $\Delta$ processes themselves are violated in cold matter, so these processes remain kinematically forbidden at the Fermi surfaces. This is because the four-fold spin degeneracy keeps the $\Delta$ Fermi momenta small, while the electron Fermi momentum is small as well. At finite temperature these processes nevertheless contribute through thermal blurring of the Fermi surfaces and collisional broadening.

\subsection{Stability of the $\Delta$ against strong decay}
\label{sec:stab}
Although the decay $\Delta\to N\pi$ remains kinematically allowed at all densities, the nucleon it would produce has a momentum well inside the nucleon Fermi sea. Each $\Delta$ species is therefore Pauli blocked against strong decay as soon as it is populated. The in-medium width is appreciable only for $\Delta^0$ and $\Delta^+$ at low density, where these species are absent or very dilute (Fig.~\ref{fig:width} in Appendix~\ref{app:width}). In the mean-field description, the populated $\Delta$s are thus stable quasiparticles whose spectral width is determined by collisions.

\subsection{Relaxation and the channels that carry it}
\begin{figure*}[t]
\includegraphics[width=\textwidth]{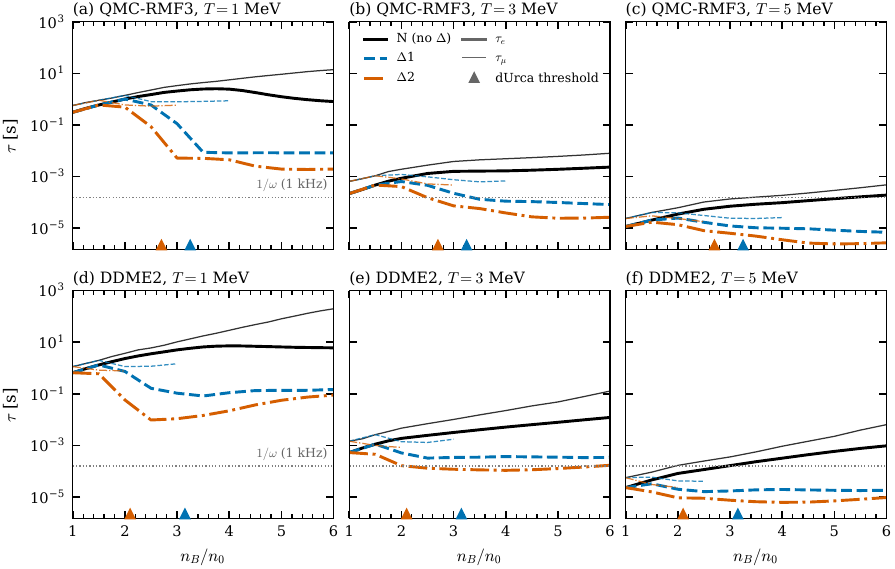}
\caption{Electron relaxation time (thick lines) and muon relaxation time (thin lines, shown where the muon fraction exceeds $10^{-3}$), Eq.~\eqref{eq:tau}, as functions of baryon density at 1, 3 and 5~MeV (columns), for QMC-RMF3 (top) and DDME2 (bottom), in isothermal, neutrino-transparent matter at the thermal weak equilibrium. Line styles distinguish the nucleonic and the two $\Delta$ sets of Table~\ref{tab:sets}; triangles on the lower axis mark the cold nucleonic direct Urca threshold of each $\Delta$ set, and the dotted grey line the inverse angular frequency at 1~kHz. The $\Delta$s shorten the electron relaxation time by one to almost three orders of magnitude above the threshold.}
\label{fig:tau}
\end{figure*}

Urca processes are the mechanism by which the lepton fractions relax to their equilibrium values. In Fig.~\ref{fig:tau} we plot the electron relaxation time with and without $\Delta$s. In nucleonic matter the direct Urca process is kinematically suppressed at low temperature, so equilibration proceeds through the collisional (``modified Urca'') contribution of the NWA; it is slow at 1~MeV and speeds up rapidly as the temperature rises. Above the $\Delta^-$ threshold the relaxation time drops sharply, by more than two orders of magnitude at 1~MeV and by one to two orders at 3--5~MeV for the sets with the larger scalar coupling. At 1~MeV the drop coincides with the opening of the nucleonic direct Urca process and appears as a step in Fig.~\ref{fig:tau}. In nucleonic matter the muon relaxation time is longer than the electron relaxation time, consistent with Ref.~\cite{AHS2023}. With $\Delta$s the muons are nearly absent above threshold, and the second mode of the coupled response plays no role.

\begin{figure*}[t]
\includegraphics[width=\textwidth]{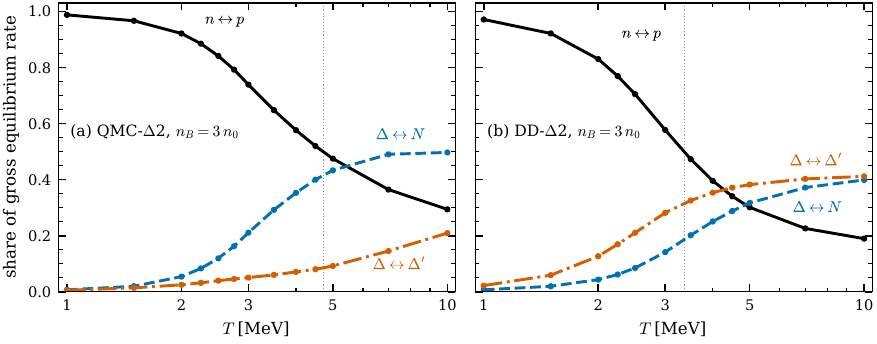}
\caption{Share of the gross equilibrium Urca rate (decay plus capture, electrons and muons) carried by the nucleonic channel, the $\Delta\leftrightarrow N$ channels ($\Delta^-\leftrightarrow n$, $\Delta^0\leftrightarrow p$) and the $\Delta\leftrightarrow\Delta'$ channels, as a function of temperature at three times $n_0$ for (a) \QDb\ and (b) \DDb; circles are the computed temperatures. The vertical dotted line marks the temperature above which the $\Delta$ channels together carry more than half of the rate. The $\Delta$ channels, closed in cold matter, take over as the temperature rises.}
\label{fig:chan}
\end{figure*}

To see which processes are responsible, in Fig.~\ref{fig:chan} we show how the equilibrium rate is shared among the channels. At 1~MeV almost all of it is carried by the nucleonic process, which the change in composition has opened. The $\Delta$ processes are forbidden in cold matter and open only through thermal blurring and collisional broadening, so their share grows with temperature; together they carry more than half of the rate above about 3.5~MeV in \DDb\ and 4.5~MeV in \QDb. Which $\Delta$ process dominates depends on the composition. In \QDb, where only the $\Delta^-$ is abundant at this density, $\Delta^-\to n$ provides most of the $\Delta$ contribution, whereas in \DDb, where the $\Delta^0$ is already present, the $\Delta\to\Delta'$ processes contribute as much or more. We see that the $\Delta$s act in two stages: at low temperature through the composition, which opens the nucleonic process, and at higher temperature through their own weak processes.

\subsection{Bulk viscosity and the resonance}
\begin{figure*}[t]
\includegraphics[width=\textwidth]{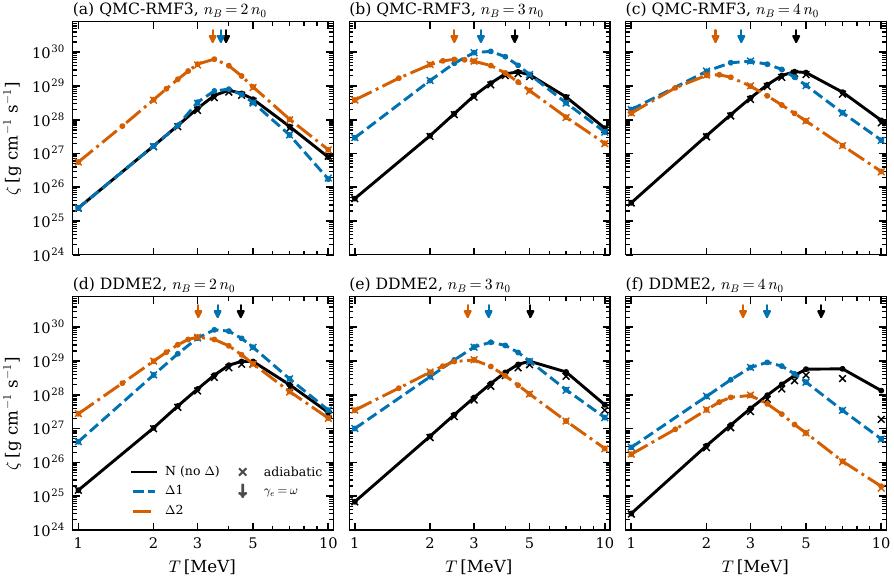}
\caption{Bulk viscosity at 1~kHz as a function of temperature at two, three and four times $n_0$ (columns) for QMC-RMF3 (top) and DDME2 (bottom). Lines with circles: isothermal oscillations at the computed temperatures; crosses: adiabatic oscillations (fixed entropy per baryon), computed on the coarser base grid. Arrows mark the resonance temperature, at which the electron relaxation rate equals the angular frequency. Line styles as in Fig.~\ref{fig:tau}. The $\Delta$s move the resonance to lower temperature, raise the bulk viscosity below the nucleonic resonance and lower it above.}
\label{fig:zetaT}
\end{figure*}

One of the physical manifestations of flavor relaxation is bulk viscosity, which damps density oscillations. In Fig.~\ref{fig:zetaT} we plot the bulk viscosity for 1~kHz oscillations as a function of temperature. The bulk viscosity attains a resonant maximum when the electron relaxation rate matches the angular frequency of the oscillation (Appendix~\ref{app:bv}). At lower temperatures the matter equilibrates so slowly that the lepton fractions are almost conserved during a cycle, and the bulk viscosity is small. At higher temperatures the matter equilibrates so quickly that there is little phase lag between pressure and density, and the bulk viscosity again becomes small.

Since the $\Delta$s speed up the relaxation, they shift the resonant peak to lower temperature, from about 4--6~MeV in nucleonic matter to about 2--3.7~MeV in the $\Delta$ sets. The shift is larger for the larger scalar coupling and at higher density, where the $\Delta^-$ fraction is larger. We obtain the resonance temperature by log--log interpolation of the electron relaxation rate between neighboring grid temperatures; removing the grid points at 2.5, 3.5, 4 and 4.5~MeV changes it by at most 0.03~MeV, so the interpolation error is negligible. The shift resembles the effect of crossing a nucleonic direct Urca threshold, which in the IUF model speeds up the relaxation above about $4\,n_0$~\cite{AHZ2024iso}. Here, however, the threshold is set by the onset of the $\Delta^-$ and therefore lies at lower density, between 1.45 and 2.3 times $n_0$ in our sets, and above about 3~MeV the $\Delta$ processes accelerate the relaxation further. The height of the peak is set by the thermodynamic strength of the response rather than by the rates, and it changes by less than about an order of magnitude in either direction. The $\Delta$s therefore have opposite effects on the two sides of the resonance. At 2--3~MeV, below the nucleonic resonance, the relaxation rate is brought closer to resonance and the bulk viscosity increases by more than an order of magnitude. Above the resonance, and in particular at four times $n_0$ and 5--10~MeV, the faster equilibration moves the rate further away from resonance, and the $\Delta$s reduce the bulk viscosity by one to almost three orders of magnitude.

At high density the composition can suppress the bulk viscosity altogether. In \DDb\ the derivative of the pressure with respect to the electron fraction and the density response of the rate change sign together near five times $n_0$, at all temperatures. According to Eq.~\eqref{eq:identity}, this is the density at which the lepton chemical potential becomes independent of density at fixed electron fraction. Compression then no longer drives the matter out of $\beta$ equilibrium, and the bulk viscosity vanishes regardless of the rates; \QDb\ approaches the same behavior at the upper end of our grid. Close to this zero, and in \DN\ at 10~MeV and high density, where the finite-temperature correction to the equilibrium condition is large, the linearized calculation yields small negative values, at 3 of the 703 points of our base grid and at 11 of the 2392 points of the denser grid used for Fig.~\ref{fig:fmaps}. Because the linearization is about a state with nonzero chemical imbalances, the density response of the rate is no longer strictly proportional to the density derivative of the imbalance, and positivity is not guaranteed. We treat these points as undamped.

Wherever the bulk viscosity is large, adiabatic and isothermal oscillations give nearly the same value (crosses in Fig.~\ref{fig:zetaT}). Appreciable differences appear only at 10~MeV and near the zero in \DDb, where the bulk viscosity is small. This is because the temperatures involved are lower than the Fermi energies of the relevant particles, so the entropy contribution to the pressure is a small correction~\cite{AHZ2024iso}. At every grid point the adiabatic modulus exceeds the isothermal one, as required by thermodynamic stability.

For nucleonic DDME2 our resonance temperature is 1--2~MeV higher than the temperature at which Ref.~\cite{AHS2023} found the electron relaxation rate to equal the oscillation frequency. In Appendix~\ref{app:ahs} we show that the rate model accounts for only a small part of this difference, and that the main source is the equilibrium state about which the rates are linearized. Reference~\cite{AHS2023} linearized about the cold condition of vanishing chemical imbalance, at which electron capture greatly exceeds neutron decay in hot matter, whereas we linearize about the finite-temperature equilibrium, following Refs.~\cite{AHHZ2021,AHZ2024iso}. When we adopt their prescription, our code reproduces their crossing temperature. The earlier study of the same model~\cite{AHS2022}, which also used the cold equilibrium, placed the 1~kHz maximum at about 5~MeV. That value refers to the total bulk viscosity of $npe\mu$ matter. For $npe\mu$ matter Ref.~\cite{AHS2023} likewise places the 1~kHz minimum of the damping time at about 5~MeV; its 3--4~MeV value refers to the crossing of the electron relaxation rate of $npe$ matter below threshold, which is the quantity we compare with. The two studies therefore do not conflict. Reference~\cite{AHS2022} also estimated that the finite-temperature correction to the equilibrium would move the maximum by about 1~MeV, which agrees with the shift we find (Appendix~\ref{app:ahs}). The $\Delta$-induced shifts discussed in this paper are differences within a single framework and do not depend on this choice.

\subsection{Frequency dependence and damping}
\begin{figure*}[t]
\includegraphics[width=\textwidth]{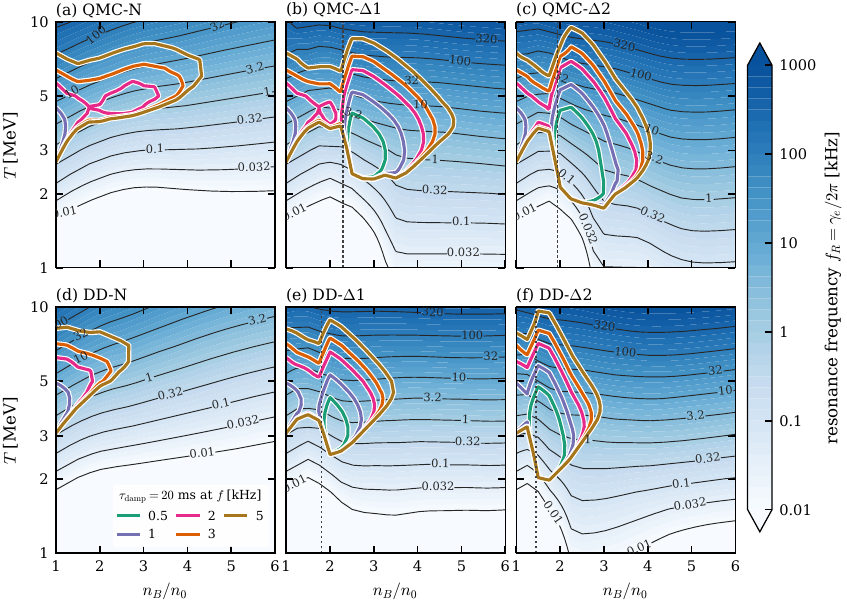}
\caption{Frequency dependence of the bulk-viscous damping in the density--temperature plane for nucleonic matter (left), \QDa\ and \DDa\ (middle) and \QDb\ and \DDb\ (right); QMC-RMF3 top, DDME2 bottom; isothermal. Blue shading and thin black contours with inline labels: resonance frequency, Eq.~\eqref{eq:fR}, in kHz. Thick colored contours, identified in the legend of panel (d): damping time of 20~ms, Eq.~\eqref{eq:tdamp}, for density oscillations at 0.5, 1, 2, 3 and 5~kHz; oscillations damp faster inside each contour. Dashed vertical lines: $\Delta^-$ onset. The isothermal grid of Table~\ref{tab:inputs} (step $0.25\,n_0$, 19 temperatures) was interpolated bilinearly in the logarithms, in density and logarithmic temperature, for display; points with non-positive bulk viscosity are treated as undamped. The $\Delta$s lower the resonance temperature at every frequency and extend the fast-damping regions to lower temperature.}
\label{fig:fmaps}
\end{figure*}

The frequency of the dominant density oscillations varies between remnants and between the oscillation modes of one remnant, so we now evaluate the response for several frequencies; the stored relaxation matrices allow this without new rate calculations. In Fig.~\ref{fig:fmaps} we show the resonance frequency, i.e., the frequency for which a given density and temperature are on resonance, together with the regions in which oscillations of 0.5 to 5~kHz are damped within 20~ms, a time comparable to the early post-merger phase.

We note the following features. (1) In the nucleonic models the resonance-frequency contours rise steeply and almost uniformly with temperature and only slowly with density, because there is no direct Urca threshold. (2) With $\Delta$s the contours bend downward above the $\Delta^-$ threshold, so that wherever $\Delta$s are present a given frequency is resonant at a temperature one to three MeV lower; at higher density they flatten or turn upward again. Doubling the frequency raises the resonance temperature by about half an MeV.

(3) The damping regions lie along the resonance contours of the corresponding frequency, since an oscillation is damped most efficiently where it is on resonance. The shortest damping time at a given frequency is then set by the thermodynamic strength of the response and scales inversely with the frequency (Appendix~\ref{app:bv}), and our results follow this scaling closely. (4) In nucleonic matter no region damps 0.5~kHz oscillations within 20~ms, 1~kHz oscillations are damped this fast only near saturation density, and the regions for higher frequencies start near 3~MeV at saturation density and move to 4--8~MeV at higher density. With $\Delta$s the damping regions extend from saturation density to three to five times $n_0$, reach down to 2--2.75~MeV and are present at every frequency we consider, with shortest damping times about 2.5 to 9 times below the nucleonic ones. At high density and temperature the damping regions close again, for the reasons discussed above.

\subsection{Sensitivity to the $\Delta$ weak inputs}
\begin{figure*}[t]
\includegraphics[width=\textwidth]{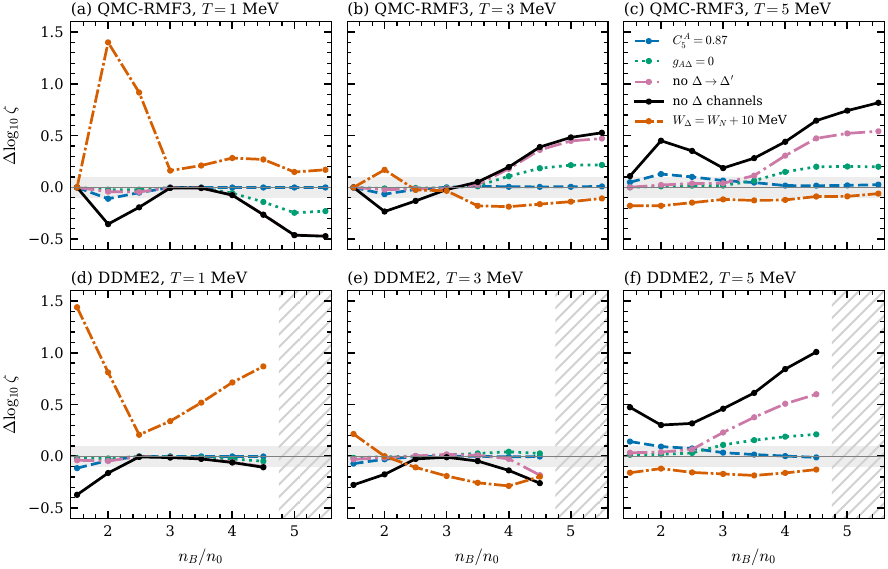}
\caption{Change of the logarithm of the bulk viscosity at 1~kHz relative to the baseline (\QDb\ and \DDb\ with the inputs of Table~\ref{tab:inputs}) when one input is varied, as a function of density at 1, 3 and 5~MeV (columns), for QMC-RMF3 (top) and DDME2 (bottom); isothermal. The grey band marks changes below 0.1~dex; the hatched region in the DDME2 panels marks densities near the zero of the bulk viscosity, where the ratio is not meaningful. The collisional width of the $\Delta$ dominates near threshold at 1~MeV, the weak couplings matter at the 0.1--0.25~dex level, and the $\Delta$ channels themselves matter most at 5~MeV and high density.}
\label{fig:sens}
\end{figure*}

Finally, in Fig.~\ref{fig:sens} we show how the bulk viscosity of \QDb\ and \DDb\ responds when single inputs are varied. The weak couplings have little influence: lowering the $N\Delta$ axial coupling or setting the $\Delta\Delta'$ axial coupling to zero changes the bulk viscosity by at most a quarter of a dex. Removing the $\Delta$ weak processes while keeping the $\Delta$s in the equation of state changes it by about half a dex or less up to 3~MeV, which confirms that the change in composition controls the response of moderately hot matter. At 5~MeV and high density, where $\Delta^0$ and $\Delta^+$ are abundant and the $\Delta$ processes carry most of the rate, the change reaches an order of magnitude. The in-medium $\Delta\to N\pi$ width has no visible effect, as expected from Pauli blocking. The results are most sensitive to the collisional width of the $\Delta$. Just above the $\Delta^-$ threshold at 1~MeV, where the $\Delta$ processes are kinematically suppressed, a larger width opens more of their phase space and increases the bulk viscosity by more than an order of magnitude, whereas at 3~MeV and above the effect stays below a third of a dex. The collisional width of the $\Delta$ in dense matter is the least constrained input of our calculation, because no microscopic calculation exists for degenerate $\Delta$s (Appendix~\ref{app:rates}).

\section{Conclusions}
\label{sec:concl}
In this work, we have calculated flavor equilibration rates and the resulting bulk viscosity of neutrino-transparent nuclear matter containing $\Delta(1232)$ isobars in strong equilibrium, at the densities and temperatures relevant to neutron star mergers. To do so we derived the spin-summed matrix elements for the charged-current $\Delta\to N$ and $\Delta\to\Delta'$ transitions from the Rarita--Schwinger spin sum, included them together with the nucleonic processes in full phase-space rates in the NWA, treated the in-medium strong width of the $\Delta$ with Pauli blocking, and obtained the bulk viscosity from the coupled response of the electron and muon fractions about the finite-temperature flavor equilibrium. The calculations were performed for two equations of state without a nucleonic direct Urca threshold, QMC-RMF3 and DDME2, each with two sets of $\Delta$ couplings.

We found that the $\Delta$s affect the bulk viscosity mainly through the composition. As the $\Delta^-$ takes over the negative charge from the leptons, the proton fraction rises and the nucleonic direct Urca process opens in models where it is otherwise closed. This controls flavor equilibration at temperatures of a few MeV, while the $\Delta$ weak processes, which are forbidden in cold matter, become dominant only at higher temperature. The faster equilibration lowers the temperature at which the bulk viscosity reaches its resonant peak, for every oscillation frequency, with little change in the height of the peak. Consistently with this, the bulk viscosity increases at 2--3~MeV, where nucleonic matter is still well below resonance, and decreases in hotter and denser matter, which is pushed further above resonance. At high density the composition can make the bulk viscosity vanish altogether. For kHz oscillations the regions of fastest damping become broader in density, extend to lower temperature and appear even below 1~kHz, with damping times several times shorter than in nucleonic matter. This can lead to effective bulk viscous damping in cooler regions of a merger if $\Delta$s are present. The populated $\Delta$s are Pauli blocked against strong decay, so they behave as long-lived quasiparticles throughout.

For nucleonic DDME2 we found that linearizing about the cold rather than the finite-temperature equilibrium lowers the resonance temperature by up to a few MeV at high density, more than any refinement of the rate model. This accounts for most of the difference between our results and those of Ref.~\cite{AHS2023}, and it should be kept in mind when different calculations are compared.

There are many directions in which further work is needed. (1) The main uncertainties lie in the $\Delta$ sector. The $\Delta$--meson couplings are not constrained by data, and we used two sets per model with a common vector coupling ratio. The collisional width of the $\Delta$ was set equal to that of the nucleon; it is the only input that changes our results by more than an order of magnitude, near the $\Delta^-$ threshold at low temperature, and a microscopic calculation of it would be valuable. The axial couplings of the $\Delta$ transitions, taken from the low-momentum-transfer form factor and from the quark model, have a minor effect. (2) Our most significant assumption about the environment is neutrino transparency, which we applied up to 10~MeV; at the higher temperatures an explicit treatment of neutrino transport will be needed. We also assumed that purely leptonic conversion between electrons and muons is frozen. (3) Our calculation assumes linear response. Large-amplitude oscillations, and coupled slow variables driven strongly, can take the matter far beyond this regime~\cite{Storbacka2026}, and a direct implementation of the rate equations in merger simulations, of the kind performed in Ref.~\cite{Most2024}, would include these effects naturally and show whether the lower resonance temperature affects the post-merger evolution and its gravitational-wave signal~\cite{Zhu2026}.

\clearpage
\appendix
\section{Relativistic mean-field model}
\label{app:rmf}
\subsection{Lagrangian}
The baryons were the nucleon doublet $\psi_N$ and the $\Delta(1232)$ quartet $\psi^\mu_\Delta$,
\begin{equation}
\psi_N=\begin{pmatrix}\psi_p\\ \psi_n\end{pmatrix},\qquad
\psi_\Delta^\mu=\begin{pmatrix}\psi^\mu_{\Delta^{++}}\\ \psi^\mu_{\Delta^{+}}\\ \psi^\mu_{\Delta^{0}}\\ \psi^\mu_{\Delta^{-}}\end{pmatrix},
\label{eq:fields}
\end{equation}
the latter described by Rarita--Schwinger (RS) vector-spinors~\cite{RS1941}. They interacted through the isoscalar-scalar $\sigma$, isoscalar-vector $\omega^\mu$ and isovector-vector $\bm\rho^\mu$ mesons. The Lagrangian density was
\begin{align}
\mathcal L ={}& \bar\psi_N\big(i\gamma_\mu D^\mu_N - m_N + g_{\sigma N}\sigma\big)\psi_N \nonumber\\
&+ \bar\psi_{\Delta\mu}\big(i\gamma_\nu D^\nu_\Delta - M_\Delta + g_{\sigma\Delta}\sigma\big)\psi^\mu_{\Delta} \nonumber\\
&+ \tfrac12\partial_\mu\sigma\partial^\mu\sigma - \tfrac12 m_\sigma^2\sigma^2 - U(\sigma) \nonumber\\
&- \tfrac14\omega_{\mu\nu}\omega^{\mu\nu} + \tfrac12 m_\omega^2\omega_\mu\omega^\mu
 - \tfrac14\bm\rho_{\mu\nu}\cdot\bm\rho^{\mu\nu} \nonumber\\
&+ \tfrac12 m_\rho^2\bm\rho_\mu\cdot\bm\rho^\mu + \mathcal L_{\omega\rho}
 + \sum_{l=e,\mu}\bar\psi_l(i\slashed\partial - m_l)\psi_l ,
\label{eq:L}
\end{align}
with the covariant derivative
\begin{equation}
D^\mu_b=\partial^\mu+i g_{\omega b}\omega^\mu + i g_{\rho b}\,\bm t_b\cdot\bm\rho^\mu ,
\label{eq:D}
\end{equation}
where $\bm t_b$ is the isospin operator of baryon $b$; its third component is $\pm\tfrac12$ for the proton and the neutron and $\tfrac32$, $\tfrac12$, $-\tfrac12$, $-\tfrac32$ for $\Delta^{++}$, $\Delta^+$, $\Delta^0$, $\Delta^-$. The RS field obeys the constraints
\begin{equation}
\gamma_\mu\psi^\mu_\Delta=0,\qquad \partial_\mu\psi^\mu_\Delta=0 ,
\label{eq:RScon}
\end{equation}
and in the mean-field approximation used below each $\Delta$ is a free spin-$\tfrac32$ quasiparticle with four spin states, an effective mass and a vector self-energy. The $\Delta$--meson couplings were written as fractions of the nucleon couplings,
\begin{equation}
g_{i\Delta}=x_{i}\,g_{iN},\qquad i=\sigma,\omega,\rho ,
\label{eq:xcoup}
\end{equation}
with the ratios $x_i$ of each coupling set listed in Table~\ref{tab:sets}.

We used two nucleonic models, whose parameters are listed in Table~\ref{tab:param}.

\textit{QMC-RMF3}~\cite{ABHT2022,ABHT2023} has constant couplings, the scalar self-interaction and the $\omega\rho$ mixing term
\begin{align}
U(\sigma)&=\tfrac{\kappa}{6}(g_\sigma\sigma)^3+\tfrac{\lambda}{24}(g_\sigma\sigma)^4 , \nonumber\\
\mathcal L_{\omega\rho}&=\Lambda_v\, g_\omega^2g_\rho^2\,(\omega_\mu\omega^\mu)(\bm\rho_\nu\cdot\bm\rho^\nu) ,
\label{eq:Uqmc}
\end{align}
with the parameters of Table~I of Ref.~\cite{ABHT2022}; Table~\ref{tab:param} gives both the original parameters and their translation into the notation of Eq.~\eqref{eq:Uqmc}. In cold $\beta$-equilibrated matter without $\Delta$s, with or without muons, the nucleonic direct Urca triangle~\cite{LPPH1991}, i.e.\ the condition that the neutron Fermi momentum does not exceed the sum of the proton and electron Fermi momenta, stays closed up to eight times $n_0$ in this model (our calculation; Ref.~\cite{Tambe2026} used QMC-RMF3 as its model without a direct Urca threshold).

\textit{DDME2}~\cite{DDME2} has no scalar self-interaction and no $\omega\rho$ term, and density-dependent couplings
\begin{align}
g_i(n_B)&=g_i(n_{\rm sat})\,f_i(x),\qquad x\equiv n_B/n_{\rm sat}, \nonumber\\
f_{\sigma,\omega}(x)&=a_i\frac{1+b_i(x+d_i)^2}{1+c_i(x+d_i)^2},\qquad f_\rho(x)=e^{-a_\rho(x-1)} ,
\label{eq:dd}
\end{align}
with the parameter values of Ref.~\cite{DDME2}. The nucleonic direct Urca triangle is also closed up to eight times $n_0$ in this model, in agreement with Refs.~\cite{Raduta2021,AHS2023}. Both models thus isolate the effect of the $\Delta$s on a background in which the nucleonic direct Urca process is closed in cold matter.

Table~\ref{tab:param} also lists the saturation properties computed with our solver. For QMC-RMF3 they agree with Ref.~\cite{ABHT2022} to the digits quoted there, with the symmetry energy and its slope defined from the difference between pure neutron matter and symmetric matter as in that reference. For DDME2 the quadratic symmetry energy and its slope agree with the values 32.3~MeV and 51.2~MeV quoted in Ref.~\cite{Raduta2021}.

\subsection{Mean-field equations and thermodynamics}
In uniform, static matter the meson fields are replaced by their expectation values, and we used the self-energies
\begin{equation}
\Phi=g_\sigma\bar\sigma,\qquad W=g_\omega\bar\omega_0,\qquad B=g_\rho\bar\rho_{0,3} .
\label{eq:selfen}
\end{equation}
Each baryon species $b$ is then a free Fermi gas with spin degeneracy $\eta_b$ (two for nucleons, four for $\Delta$s), effective mass and single-particle energy
\begin{align}
m^*_b &= M_b - x_{\sigma b}\Phi , \label{eq:mstar}\\
E_b(\bm k) &= \sqrt{k^2+m_b^{*2}} + U_b, \nonumber\\
U_b &= x_{\omega b}W + x_{\rho b}\,\theta_b\,B + \Sigma_R , \label{eq:Ub}
\end{align}
where the coupling ratios are one for nucleons, the isospin factor $\theta_b$ is given in Table~\ref{tab:param}, and the rearrangement self-energy $\Sigma_R$ of Eq.~\eqref{eq:SigR} is present only for density-dependent couplings. With the effective chemical potential, the kinetic energy and the occupation
\begin{align}
\nu_b&\equiv\mu_b-U_b,\qquad E^*_k=\sqrt{k^2+m^{*2}_b}, \nonumber\\
f_b&=\frac{1}{1+e^{(E^*_k-\nu_b)/T}} ,
\label{eq:occ}
\end{align}
the kinetic contributions of species $b$ at temperature $T$ are
\begin{align}
n_b &= \eta_b\!\int\!\frac{d^3k}{(2\pi)^3} f_b ,\qquad
n^s_b = \eta_b\!\int\!\frac{d^3k}{(2\pi)^3}\frac{m^*_b}{E^*_k} f_b , \nonumber\\
P^{\rm kin}_b &= \frac{\eta_b}{3}\!\int\!\frac{d^3k}{(2\pi)^3}\frac{k^2}{E^*_k} f_b , \nonumber\\
\varepsilon^{\rm kin}_b &= \eta_b\!\int\!\frac{d^3k}{(2\pi)^3}E^*_k f_b , \nonumber\\
s_b &= \big(\varepsilon^{\rm kin}_b + P^{\rm kin}_b - \nu_b n_b\big)/T .
\label{eq:fg}
\end{align}
Antiparticles were neglected. The most exposed case is the electron at the smallest electron fraction we encountered, about $2\times10^{-3}$ at three times $n_0$, where the electron chemical potential is about 70~MeV; at 10~MeV the positron density is then suppressed by a factor of about $10^{-3}$.

The field equations follow from the stationarity of the grand potential $\Omega$ with respect to $\Phi$, $W$ and $B$. For QMC-RMF3,
\begin{align}
\frac{m_\sigma^2}{g_\sigma^2}\Phi + \frac{\kappa}{2}\Phi^2 + \frac{\lambda}{6}\Phi^3 &= \sum_b x_{\sigma b}n^s_b , \nonumber\\
\frac{m_\omega^2}{g_\omega^2}W + 2\Lambda_v B^2 W &= \sum_b x_{\omega b}n_b , \nonumber\\
\Big(\frac{m_\rho^2}{g_\rho^2} + 2\Lambda_v W^2\Big)B &= \sum_b x_{\rho b}t_{3b}n_b ,
\label{eq:fieldQMC}
\end{align}
and the pressure is
\begin{align}
P ={}& \sum_b P^{\rm kin}_b - \frac{m_\sigma^2}{2g_\sigma^2}\Phi^2 - \frac{\kappa}{6}\Phi^3 - \frac{\lambda}{24}\Phi^4 \nonumber\\
&+ \frac{m_\omega^2}{2g_\omega^2}W^2 + \frac{m_\rho^2}{2g_\rho^2}B^2 + \Lambda_v W^2B^2 + P_\ell ,
\label{eq:PQMC}
\end{align}
where $P_\ell$ is the lepton pressure. For DDME2 the couplings depend on the baryon density, so the derivative of the grand potential with respect to the baryon densities acquires the rearrangement term
\begin{align}
\Sigma_R = \sum_b\Big[&\frac{\partial\ln g_\omega}{\partial n_B}x_{\omega b}W n_b + \frac{\partial\ln g_\rho}{\partial n_B}x_{\rho b}2t_{3b}B n_b \nonumber\\
&- \frac{\partial\ln g_\sigma}{\partial n_B}x_{\sigma b}\Phi n^s_b\Big] ,
\label{eq:SigR}
\end{align}
which enters every $U_b$ in Eq.~\eqref{eq:Ub}. The DDME2 field equations and pressure are
\begin{align}
\Phi&=\frac{g_\sigma^2}{m_\sigma^2}\sum_b x_{\sigma b}n^s_b,\qquad
W=\frac{g_\omega^2}{m_\omega^2}\sum_bx_{\omega b}n_b, \nonumber\\
B&=\frac{g_\rho^2}{m_\rho^2}\sum_b x_{\rho b}2t_{3b}n_b, \nonumber\\
P&=\sum_b P^{\rm kin}_b - \frac{m_\sigma^2\Phi^2}{2g_\sigma^2} + \frac{m_\omega^2W^2}{2g_\omega^2} \nonumber\\
&\quad+ \frac{m_\rho^2B^2}{2g_\rho^2} + n_B\Sigma_R + P_\ell .
\label{eq:PDD}
\end{align}
The term containing the rearrangement self-energy is required for thermodynamic consistency: it makes the pressure obtained from the grand potential coincide with the pressure obtained from the derivative of the energy per baryon at fixed composition and entropy per baryon. In both models the energy density is
\begin{equation}
\varepsilon=-P+Ts+\sum_i\mu_i n_i .
\label{eq:eps}
\end{equation}

\subsection{Stability and consistency}
(ii) \textit{Spinodal stability.} We computed the curvature matrix of the free energy at fixed temperature in the variables baryon density and lepton density,
\begin{equation}
C=\frac{\partial(\mu_B,\mu_L)}{\partial(n_B,n_L)},\qquad \mu_L=\mu_Q+\mu_e .
\label{eq:spin}
\end{equation}
It was positive definite between one and six times $n_0$, at 1 and 5~MeV, for all four $\Delta$ sets; the smallest eigenvalue, 144~MeV\,fm$^3$, occurred for \DDb\ at 1.5 times $n_0$. This is consistent with Ref.~\cite{Raduta2021}, which found $\Delta$-driven instabilities in DDME2 only when the scalar ratio exceeds the vector ratio by 0.15 to 0.2; in our sets the excess is at most 0.1. The numerical matrix was symmetric, as the Maxwell relation requires, to $7\times10^{-3}$ or better.

The energy density of Eq.~\eqref{eq:eps}, evaluated with an independently computed pressure, was consistent to a relative accuracy of $8\times10^{-9}$.

\section{Urca rates: phase space, nucleon width approximation and nucleon matrix element}
\label{app:rates}
\subsection{Phase-space integral}
For a decay of baryon 1 into baryon 2, a charged lepton and an antineutrino, with mean-field single-particle energies $E_i$ (kinetic energy $E_i^*$ plus vector self-energy), the rate is
\begin{align}
\Gamma^{\rm dec} ={}& \int\!\frac{d^3p_1}{(2\pi)^32E_1^*}\frac{d^3p_2}{(2\pi)^32E_2^*} \nonumber\\
&\times\frac{d^3k_l}{(2\pi)^32E_l}\frac{d^3k_\nu}{(2\pi)^32E_\nu} \nonumber\\
&\times (2\pi)^4\delta^4(P_1-P_2-k_l-k_\nu) \nonumber\\
&\times \overline{\sum}|\mathcal M|^2\, f_1(1-f_2)(1-f_l) ,
\label{eq:rate12}
\end{align}
where the four-vectors $P_i$ carry the canonical energies $E_i$ and the momenta $\bm p_i$, and $\overline\sum|\mathcal M|^2$ denotes the sum over the spins of all four particles. No average over the initial spin is taken, because the occupation $f_1$ already counts each spin state. The spinor normalization to twice the effective mass makes the kinetic energies $E^*_i$ appear in the flux factors, while the canonical energies enter energy conservation, because the vector mean fields shift the energies but not the Dirac structure. The capture rate follows from Eq.~\eqref{eq:rate12} by replacing the statistical factor by $(1-f_1)f_2f_l$ and the outgoing antineutrino by an incoming lepton and outgoing neutrino.

We reduced Eq.~\eqref{eq:rate12} exactly to five dimensions. The neutrino momentum integral removes the momentum delta function through
\begin{equation}
\int\frac{d^3k_\nu}{2E_\nu}=\int d^4k_\nu\,\delta(k_\nu^2)\,\theta(k_\nu^0) ,
\label{eq:nuint}
\end{equation}
which turns energy conservation into
\begin{equation}
\delta\big(E_\nu^2-|\bm p_1-\bm p_2-\bm k_l|^2\big),\qquad E_\nu\equiv E_1-E_2-E_l .
\label{eq:econs}
\end{equation}
We chose the momentum of baryon 1 along the $z$ axis, so that its solid angle gives $4\pi$, put the momentum of baryon 2 in the $xz$ plane at polar cosine $z_2$, so that its azimuth gives $2\pi$, and gave the lepton momentum polar cosine $z_l$ and azimuth $\phi_l$. With
\begin{equation}
d^3p=p\,E^*\,dE^*\,d\Omega
\label{eq:d3p}
\end{equation}
for each massive particle, and trading the energy of baryon 1 for the neutrino energy at fixed baryon-2 and lepton energies,
\begin{align}
|\bm p_1-\bm p_2-\bm k_l|^2 &= R(z_2,z_l) - S(z_2,z_l)\cos\phi_l , \nonumber\\
R&=p_1^2+p_2^2+k_l^2-2p_1k_lz_l \nonumber\\
&\quad-2p_1p_2z_2+2p_2k_lz_2z_l , \nonumber\\
S&=2p_2k_l\sqrt{1-z_2^2}\sqrt{1-z_l^2} .
\label{eq:RS}
\end{align}
The azimuthal integral has two roots, each with Jacobian
\begin{equation}
|S\sin\phi_l|=\sqrt{S^2-(R-E_\nu^2)^2} .
\label{eq:jac}
\end{equation}
Collecting factors,
\begin{align}
\Gamma^{\rm dec} ={}& \frac{1}{128\pi^6}\int\! dE_2\,dE_l\,dE_\nu\,dz_2\,dz_l\,
\frac{p_1p_2k_l\,\overline\sum|\mathcal M|^2}{\sqrt{S^2-(R-E_\nu^2)^2}} \nonumber\\
&\times f_1(1-f_2)(1-f_l)\big|_{E_1=E_2+E_l+E_\nu} .
\label{eq:rate5}
\end{align}
For fixed $z_2$ the radicand is a quadratic in $z_l$ with two roots given in closed form by the kinematics, so the $z_l$ integral of the polynomial numerator is evaluated exactly by Gauss--Chebyshev quadrature. The remaining four integrals were evaluated by scrambled Sobol quasi-Monte Carlo over energy windows around the Fermi energies; the window width, the number of points and the number of Chebyshev nodes are given in Table~\ref{tab:inputs}. The allowed range of $z_2$ is the one for which a real azimuth exists. This reduction is the one of Ref.~\cite{AHHZ2021}, extended here to unequal baryon masses, arbitrary vector potentials and spin-$\tfrac32$ baryons, which enter only through the matrix element.

\subsection{Nucleon width approximation}
Above about 1~MeV the collisional broadening of the baryon quasiparticles opens the reactions below the direct-Urca threshold that are usually treated as modified Urca. We used the nucleon width approximation (NWA) of Ref.~\cite{AHZ2024nwa}: each baryon of the Urca reaction is given a Breit--Wigner spectral function in its mass,
\begin{equation}
\mathcal A_b(m) = \frac{1}{\pi}\frac{W_b/2}{(m-m^*_b)^2+W_b^2/4} ,
\label{eq:BW}
\end{equation}
and the rate is the spectral average
\begin{equation}
\Gamma_{\rm NWA} = \int dm_1\,dm_2\;\mathcal A_1(m_1)\mathcal A_2(m_2)\,\Gamma(m_1,m_2) ,
\label{eq:nwa}
\end{equation}
with $\Gamma(m_1,m_2)$ the rate of Eq.~\eqref{eq:rate5} evaluated with masses $m_1$ and $m_2$ in the dispersion relations and the matrix element. Following Refs.~\cite{AHZ2024nwa,Tambe2026}, the nucleon width grows quadratically with temperature,
\begin{equation}
W_N=\frac{T^2}{T_W},
\label{eq:WN}
\end{equation}
with the scale $T_W$ of Table~\ref{tab:inputs}. For the $\Delta$s we added the in-medium $\Delta\to N\pi$ width of Sec.~\ref{sec:width},
\begin{equation}
W_\Delta=W_N+\Gamma^{\rm med}_\Delta ,
\label{eq:WD}
\end{equation}
so the collisional broadening of the $\Delta$ was set equal to that of the nucleon. Calculations of the in-medium $\Delta$ width concern $\Delta$s excited above the nucleon Fermi sea, in nuclei and in hot hadronic matter, where the $\Delta\to N\pi$ channel and its medium modification dominate~\cite{vanHeesRapp2004}. We are not aware of a calculation of the collisional width of degenerate $\Delta$s near their own Fermi surface. For any degenerate fermion this width vanishes at the Fermi surface as $T^2$, which is the form of Eq.~\eqref{eq:WN}; its coefficient for the $\Delta$ is not known, and in Fig.~\ref{fig:sens} we test it by adding 10~MeV to $W_\Delta$. We sampled Eq.~\eqref{eq:nwa} jointly with the kinematics through the substitution
\begin{equation}
m=m^*+\tfrac12W\sinh t,\qquad \mathcal A(m)\,dm=\frac{dt}{\pi\cosh t} ,
\label{eq:sinh}
\end{equation}
mixed with a uniform component to cover the tails. The resulting six-dimensional quasi-Monte Carlo integral agreed with the nested evaluation of Eq.~\eqref{eq:nwa} to 0.015~dex.

\subsection{Lepton tensor and nucleon matrix element}
The effective charged-current interaction and the lepton current are
\begin{align}
\mathcal L_W&=-\frac{G}{\sqrt2}J^\mu_{\rm had}\,l_\mu+{\rm h.c.},\qquad G=G_F\cos\theta_C, \nonumber\\
l_\mu&=\bar u_l\gamma_\mu(1-\gamma_5)v_\nu ,
\label{eq:LW}
\end{align}
with the constants of Table~\ref{tab:inputs}. The lepton tensor, summed over spins, is
\begin{align}
L_{\mu\nu}&=\mathrm{Tr}\big[\slashed k\gamma_\mu(1-\gamma_5)\slashed k'\gamma_\nu(1-\gamma_5)\big] \nonumber\\
&=8\big[k_{\mu}k'_{\nu}+k_{\nu}k'_{\mu}-g_{\mu\nu}\,k\!\cdot\!k' - i\epsilon_{\mu\nu\alpha\beta}k^\alpha k'^\beta\big] ,
\label{eq:Ltensor}
\end{align}
where $k$ and $k'$ are the charged-lepton and neutrino momenta; the lepton mass drops out because of the chiral projectors.

\textit{Nucleons.} With the nucleon current
\begin{equation}
J^\mu=\bar u_p\gamma^\mu(1-g_A\gamma_5)u_n ,
\label{eq:JN}
\end{equation}
the standard trace gives
\begin{align}
\overline\sum|\mathcal M|^2 = 32G^2\big[&(1+g_A)^2(p_p\!\cdot\!k_l)(p_n\!\cdot\!k_\nu) \nonumber\\
&+ (1-g_A)^2(p_n\!\cdot\!k_l)(p_p\!\cdot\!k_\nu) \nonumber\\
&+ (g_A^2-1)m_n^*m_p^*\,(k_l\!\cdot\!k_\nu)\big] ,
\label{eq:MN}
\end{align}
which for nucleons at rest averages over lepton angles to
\begin{equation}
\overline\sum|\mathcal M|^2\to32G^2(1+3g_A^2)\,m^*_nm^*_p\,E_lE_\nu .
\label{eq:MNnr}
\end{equation}

\section{$\Delta\to\Delta'$ spin sum}
\label{app:DD}
With the current and isospin factor of Eq.~\eqref{eq:JDD}, the spin-summed squared amplitude is
\begin{align}
\overline\sum|\mathcal M|^2&=\frac{G^2}{2}|\langle T_+\rangle|^2\,\mathcal H, \nonumber\\
\mathcal H&=g_{\alpha\alpha'}g_{\beta\beta'}\,\mathrm{Tr}\big[\Gamma^\mu P_1^{\alpha'\beta'}\bar\Gamma^\nu P_2^{\beta\alpha}\big]L_{\mu\nu}, \nonumber\\
\Gamma^\mu&=\gamma^\mu(1-g_{A\Delta}\gamma_5),\qquad \bar\Gamma^\nu=\gamma^0\Gamma^{\nu\dagger}\gamma^0 ,
\label{eq:Hdefdd}
\end{align}
where the $P_i$ are the RS spin sums of Eq.~\eqref{eq:RSsum} for the two $\Delta$s. In terms of the invariants
\begin{align}
a&=p_1\!\cdot\!p_2,\quad b=p_1\!\cdot\!k_l,\quad c=p_1\!\cdot\!k_\nu, \nonumber\\
d&=p_2\!\cdot\!k_l,\quad e=p_2\!\cdot\!k_\nu,\quad f=k_l\!\cdot\!k_\nu ,
\label{eq:inv}
\end{align}
and the effective masses $M_1$, $M_2$ of the two $\Delta$s, FORM gave
\begin{align}
\mathcal H ={}& \mathcal H_0+g_{A\Delta}\mathcal H_1+g_{A\Delta}^2\mathcal H_2, \nonumber\\
\mathcal H_0 ={}& \tfrac{256}{9}\Big[-af+\tfrac{a^2(be+cd)}{M_1^2M_2^2}-\tfrac{a^2f}{M_1M_2}+\tfrac{bcM_2}{M_1}+\tfrac{deM_1}{M_2}\Big] \nonumber\\
&+\tfrac{640}{9}\big(be+cd-fM_1M_2\big) , \nonumber\\
\mathcal H_1 ={}& \tfrac{512}{9}\,\tfrac{a^2(cd-be)}{M_1^2M_2^2}+\tfrac{256}{3}(cd-be) , \nonumber\\
\mathcal H_2 ={}& \tfrac{256}{9}\Big[-af+\tfrac{a^2(be+cd)}{M_1^2M_2^2}+\tfrac{a^2f}{M_1M_2}-\tfrac{bcM_2}{M_1}-\tfrac{deM_1}{M_2}\Big] \nonumber\\
&+\tfrac{640}{9}\big(be+cd+fM_1M_2\big) .
\label{eq:HDD}
\end{align}
The term odd in the axial coupling is the vector--axial interference, which changes sign between decay and capture because the roles of the charged lepton and the neutrino are exchanged. For both $\Delta$s at rest with equal masses, and with $\theta$ the angle between the lepton momenta,
\begin{equation}
\mathcal H\to\frac{128}{9}M^2E_lE_\nu\big[9(1+\cos\theta)+5g_{A\Delta}^2(3-\cos\theta)\big] ,
\label{eq:Hrest}
\end{equation}
whose angular average gives Eq.~\eqref{eq:MDDnr}.
\section{One-variable limit, oscillation type, damping time and resonance}
\label{app:bv}
Equation~\eqref{eq:Gmat} was evaluated at fixed temperature (isothermal) and at fixed entropy per baryon (adiabatic). In the adiabatic case every perturbed state was assigned the temperature at which the entropy per baryon equals its equilibrium value, found by root finding, and the rates were re-evaluated at that temperature. The adiabatic modulus exceeded the isothermal one at every grid point, by at most 12\%, as required by thermodynamic stability.

The energy stored in a density oscillation and the mean dissipated power per unit volume are
\begin{align}
\mathcal E&=\tfrac12\kappa^{-1}\Big(\frac{\delta n}{n_B}\Big)^2,\qquad
\dot{\mathcal E}=\tfrac12\zeta\omega^2\Big(\frac{\delta n}{n_B}\Big)^2, \nonumber\\
\kappa^{-1}&=n_B\frac{\partial P}{\partial n_B}\Big|_{Y_e,Y_\mu} ,
\label{eq:energy}
\end{align}
so the damping time of the amplitude is
\begin{equation}
\tau_{\rm damp}=\frac{\kappa^{-1}}{\omega^2\zeta} ,
\label{eq:tdamp}
\end{equation}
equivalent to the form in terms of the incompressibility used in Ref.~\cite{AHS2023}. The oscillation frequency is given in Table~\ref{tab:inputs}.

For a single slow variable, Eq.~\eqref{eq:zeta1} can be written in the resonance form
\begin{equation}
\zeta=\mathcal C\,\frac{\gamma}{\gamma^2+\omega^2},\qquad
\mathcal C\equiv-\frac{\partial P}{\partial Y_e}\,\frac{\gamma_B}{\gamma} .
\label{eq:resform}
\end{equation}
To the extent that the net rate depends on the state only through the chemical imbalance, the ratio of the density response to the relaxation rate in the strength $\mathcal C$ is a ratio of thermodynamic derivatives, and $\mathcal C$ does not depend on the magnitude of the rates. At fixed frequency the bulk viscosity is then largest on resonance,
\begin{equation}
\gamma=\omega:\qquad \zeta_{\max}=\frac{\mathcal C}{2\omega},\qquad
\tau_{\rm damp}^{\min}=\frac{2\kappa^{-1}}{\omega\,\mathcal C} ,
\label{eq:res}
\end{equation}
so the height of the peak is set by thermodynamics alone and the shortest damping time at a given frequency is inversely proportional to the frequency. We define the resonance frequency of a state as
\begin{equation}
f_R=\frac{\mathcal G_{ee}}{2\pi} .
\label{eq:fR}
\end{equation}

\section{In-medium $\Delta$ width}
\label{app:width}
\begin{figure*}[t]
\includegraphics[width=\textwidth]{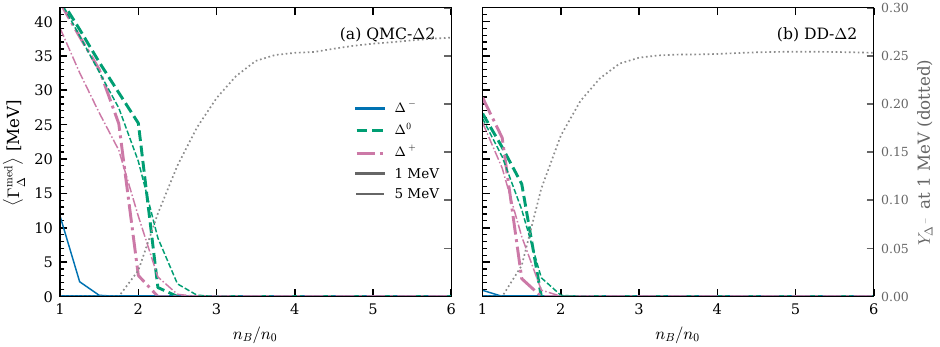}
\caption{Fermi-surface averaged in-medium $\Delta\to N\pi$ width, Eq.~\eqref{eq:Gmed}, for $\Delta^-$, $\Delta^0$ and $\Delta^+$ in $\beta$-equilibrated matter for (a) \QDb\ and (b) \DDb, at 1~MeV (thick) and 5~MeV (thin). Dotted grey line, right axis: $\Delta^-$ fraction at 1~MeV. The decay is Pauli blocked wherever the decaying species is populated; the nonzero widths at low density belong to species that are absent there.}
\label{fig:width}
\end{figure*}

Figure~\ref{fig:width} shows the Fermi-surface averaged width of Eq.~\eqref{eq:Gmed} discussed in Sec.~\ref{sec:stab}.

\section{Comparison with Ref.~\cite{AHS2023}}
\label{app:ahs}
To locate the difference between our nucleonic DDME2 resonance temperatures and those of Ref.~\cite{AHS2023}, we evaluated the electron relaxation rate of Eq.~\eqref{eq:Gmat} in matter without muons in four ways, at one, three and five times the DDME2 saturation density used there, which we denote $n_0'$ (0.152~fm$^{-3}$). This rate is the same quantity as the product of the rate coefficient and the susceptibility used in Ref.~\cite{AHS2023},
\begin{equation}
\mathcal G_{ee}=\lambda_eA_1,\qquad \frac{\partial\,\delta\mu_e}{\partial Y_e}\Big|_{n_B}=-n_BA_1 .
\label{eq:ahsdef}
\end{equation}
The four variants were: A, the prescription of Ref.~\cite{AHS2023}, namely direct Urca rates without width, only the first term of Eq.~\eqref{eq:MN}, and the derivative taken at the cold equilibrium, where the imbalance of Eq.~\eqref{eq:dmu} vanishes; B, our NWA rates with the full matrix element, at the cold equilibrium; C, our NWA rates at the thermal equilibrium of Eq.~\eqref{eq:thermeq}, which is what the main text uses; D, direct Urca without width at the thermal equilibrium.

\begin{figure*}[t]
\includegraphics[width=\textwidth]{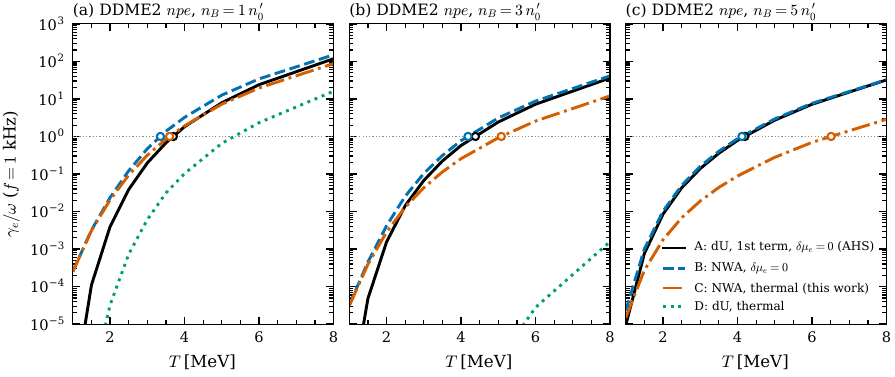}
\caption{Electron relaxation rate in units of the angular frequency at 1~kHz for nucleonic DDME2 matter without muons at (a) one, (b) three and (c) five times $n_0'$ (0.152~fm$^{-3}$). A (solid black): the prescription of Ref.~\cite{AHS2023} reproduced with our code (direct Urca without width, first term of the matrix element, linearized about the cold equilibrium). B (dashed blue): NWA rates with the full matrix element, same linearization point. C (dash-dotted orange): NWA linearized about the thermal equilibrium, as in the main text. D (dotted green): direct Urca linearized about the thermal equilibrium. Open circles mark the crossings with the dotted grey line, which move mainly with the linearization point, not with the rate model.}
\label{fig:ahs}
\end{figure*}

Figure~\ref{fig:ahs} shows the four relaxation rates in units of the angular frequency. Variant A crossed the oscillation frequency at 3.7, 4.4 and 4.2~MeV, which agrees to within 0.4~MeV with the statement of Ref.~\cite{AHS2023} that the crossing lies at 3--4~MeV below threshold. The remaining difference is of the size expected from the normalization of the coupling: Ref.~\cite{AHS2023} quotes the combination
\begin{equation}
G=G_F\cos\theta_c\,(1+g_A^2)
\label{eq:ahsG}
\end{equation}
in its reduced rate formula, and a change of the rate by the corresponding factor of 1.31 relative to the first-term normalization would lower the crossing by about 0.2~MeV. Our variant A also reproduced two qualitative statements of Ref.~\cite{AHS2023}: at the cold equilibrium the electron-capture rate exceeded the neutron-decay rate by more than three orders of magnitude, and at 5~MeV it depended only weakly on density (2.7, 1.6 and $2.5\times10^{-13}$~MeV$^4$ at the three densities).

Variant B differs from A only in the rate model and crossed at 3.3, 4.2 and 4.1~MeV, whereas variant C, which differs from B only in the linearization point, crossed at 3.6, 5.1 and 6.5~MeV. The rate model thus accounts for at most 0.4~MeV of the difference and the linearization point for 0.3--2.4~MeV, the latter growing with density. For direct Urca alone (variant D) the thermal equilibrium lies at an electron imbalance of about 20--160~MeV, and above $n_0'$ the relaxation rate remains far below the oscillation frequency up to 8~MeV. Below threshold, direct Urca without collisional broadening therefore cannot describe equilibration about the thermal equilibrium, which is one reason the NWA is required.

At the cold equilibrium the net rate does not vanish, so the background about which variant A is linearized is not stationary; Ref.~\cite{AHHZ2021} showed that using this naive condition at 3~MeV overestimates electron capture by an order of magnitude. For this reason we used the thermal equilibrium throughout, following Refs.~\cite{AHHZ2021,AHZ2024iso,Tambe2026}. For the same model, Ref.~\cite{AHS2022} reported a 1~kHz resonant maximum at about 5~MeV, also with the cold equilibrium condition, and estimated that the finite-temperature correction would move the maximum by about 1~MeV. That maximum is the one of the total $npe\mu$ bulk viscosity. For $npe\mu$ matter Ref.~\cite{AHS2023} likewise places the 1~kHz minimum of the damping time at about 5~MeV, whereas its 3--4~MeV value is the crossing of the electron relaxation rate of $npe$ matter below threshold. The two published values therefore describe different quantities and do not conflict. The shift we find between variants B and C at three times $n_0'$, 0.9~MeV, agrees with the estimate of Ref.~\cite{AHS2022}.

\bibliography{refs}

\end{document}